\documentclass[multphys,vecphys]{svmult}

\usepackage{makeidx}     
\usepackage{graphicx}    
\usepackage{multicol}    

\makeindex             

\begin{document}

\title*{Inflationary Cosmological Perturbations of Quantum-Mechanical
Origin}
\author{J\'er\^ome Martin\inst{1}}
\institute{Institut d'Astrophysique de
Paris, GreCO, FRE 2435-CNRS, 98bis boulevard Arago, 75014 Paris,
France \texttt{jmartin@iap.fr}}
%
%
\maketitle

This review article aims at presenting the theory of inflation. We
first describe the background spacetime behavior during the slow-roll
phase and analyze how inflation ends and the Universe reheats. Then,
we present the theory of cosmological perturbations with special
emphasis on their behavior during inflation. In particular, we discuss
the quantum-mechanical nature of the fluctuations and show how the
uncertainty principle fixes the amplitude of the perturbations. In a
next step, we calculate the inflationary power spectra in the
slow-roll approximation and compare these theoretical predictions to
the recent high accuracy measurements of the Cosmic Microwave
Background radiation (CMBR) anisotropy. We show how these data already
constrain the underlying inflationary high energy physics. Finally, we
conclude with some speculations about the trans-Planckian problem,
arguing that this issue could allow us to open a window on physical
phenomena which have never been probed so far.

\section{Introduction}
\label{sec:1}
Inflation is the most promising theory of the early Universe. It was
invented by A.~Guth~\cite{guth} at the beginning of the $80$'s in
order to solve the puzzles of the hot Big Bang theory. A very
interesting aspect of the inflationary theory is that it allows us to
build a bridge between cosmology and high energy physics. This is
particularly valuable in view of the fact that it is difficult to
probe physics beyond the standard model of particular physics.

\par

However, the details of the underlying particle physics model are
encoded into the fine structure of the cosmological observables. This
is why, after the invention of the inflationary scenario and during
quite a long time, it was in fact only possible to check the
consistency of the inflationary predictions. The situation has now
changed drastically with the recent releases of very high accuracy
cosmological data. One can now take advantage of the full predictive
power of inflation with the hope to learn about physics in a regime
which has never been reached before.

\par

The goal of this review article is to give a general presentation of
the inflationary scenario. In particular, we will emphasize how the
origin of the inhomogeneities present in our Universe is explained in
the framework of inflation. We will see that it is based on an elegant
interplay between general relativity and quantum theory. Then, we will
present the corresponding predictions made by inflation and will study
how the currently available data can already put some constraints on
the underlying particle physics models.

\par

This article is organized as follows. In the next section, we describe
the evolution of the inflationary background, the slow-roll phase and
the reheating. Then, we present the theory of cosmological
perturbations of quantum-mechanical origin and compare its predictions
to the available data. Finally, we conclude this article with some
speculations concerning the trans-Planckian problem of inflation,
demonstrating that future astrophysical observations will maybe open a
new window on high energy physics.

\section{The Inflationary Universe}
\label{sec:2}

\subsection{Basic Equations}

The cosmological principle implies that the Universe is, on large
scales, homogeneous and isotropic. As a consequence, the metric tensor
which describes the geometry of the Universe is of the
Friedman-Lema\^{\i}tre-Robertson-Walker (FLRW) form, namely
\begin{equation}
\label{metric0}
{\rm d}s^2=-c^2{\rm d}t^2+a^2(t)\gamma _{ij}^{(3)}{\rm d}x^i{\rm d}x^j
=a^2(\eta )\left[-{\rm d}\eta ^2+\gamma _{ij}^{(3)}{\rm d}x^i{\rm
d}x^j \right]\, ,
\end{equation}
where $\gamma _{ij}^{(3)}$ is the metric of the three-dimensional
spacelike sections. The three-dimensional sections have a constant
scalar curvature. The variable $t$ is the cosmic time while $\eta $ is
the conformal time. They are linked by the relation $c{\rm d}t=a{\rm
d}\eta $. In this article, we will work with dimensionless coordinates
$x^i$ and, as a consequence, the scale factor $a(\eta )$ will have the
dimension of a length.

\par

The matter is assumed to be a collection of $N$ perfect fluids and
therefore its stress-energy tensor is given by the following
expression
\begin{equation}
T_{\mu \nu }=\sum _{i=1}^NT_{\mu \nu}^{(i)}=(\rho _{_{\rm T}}+p_{_{\rm
T}})u_{\mu }u_{\nu}+p_{_{\rm T}}g_{\mu \nu}\, ,
\end{equation}
where $\rho _{_{\rm T}}$ is the (total) energy density and $p_{_{\rm
T}}$ the (total) pressure. These two quantities are linked by the
equation of state, $p_{_{\rm T}}=\omega \left(\rho _{_{\rm T}}\right)$
[in general, there is an equation of state per fluid considered,
i.e. $p_i=\omega _i(\rho _i)$]. The vector $u_{\mu }$ is the four
velocity and satisfy the relation $u_{\mu }u^{\mu }=-1$. This means
that one has $u^{\mu }=(1/a,0)$ and $u_{\mu }=(-a,0)$. The fact that
the stress-energy tensor is conserved, $\nabla ^{\alpha }T_{\alpha
\mu}=0$, amounts to
\begin{equation}
\label{conservation}
\rho _{_{\rm T}}'+3\frac{a'}{a}(\rho _{_{\rm T}}+p_{_{\rm T}})=0\, .
\end{equation}
This expression is obtained from the $\mu =0$ component. The component
$\mu =i$ does not lead to an interesting equation for the
background. If one assumes that each fluid is separately conserved
then the above equation is valid for each species.

\par

We will assume that gravity is correctly described by the theory of
General Relativity even in the very early Universe. This means that
the equations which link the geometrical part to the matter part are
nothing but the Einstein equations
\begin{equation}
R_{\mu \nu}-\frac{1}{2}Rg_{\mu \nu}=\kappa T_{\mu \nu}\, ,
\end{equation}
where $\kappa \equiv 8\pi G/c^4=8\pi /m_{_{\rm Pl}}^2$. These
equations, in the case of a FLRW Universe, are differential equations
determining the time evolution of the scale factor and read
\begin{equation}
\frac{3}{a^2}\biggl[\biggl(\frac{a'}{a}\biggr)^2+k\biggr]=\kappa 
\sum _{i=1}^{N}\rho _i, \quad -\frac{1}{a^2}\biggr[2\frac{a''}{a}
-\biggl(\frac{a'}{a}\biggr)^2+k\biggr]=\kappa 
\sum _{i=1}^{N}p_i\, ,
\end{equation}
where a prime denotes a derivative with respect to conformal time. In
the following, we will use the definition ${\cal H}\equiv a'/a$. The
parameter $k=0,\pm 1$ represents the curvature of the spacelike
sections. If, in addition, the equation of state of the perfect fluids
are provided, then we have a closed system of differential equations
and, therefore, the evolution of the corresponding model of the
Universe is completely specified.

\subsection{The Inflationary Hypothesis}
\index{inflation!definition}

By definition, inflation is a phase of accelerated expansion, i.e. the
scale factor satisfies~\cite{inflation}
\begin{equation}
\label{definf}
\frac{{\rm d}^2a}{{\rm d}t^2}>0\, .
\end{equation}
It is interesting to postulate that such a phase took place in the
very early Universe because, in this case, one can explain many
different seemingly paradoxical facts like, for instance, the horizon
problem or the flatness problem. Because of the latter, from now on,
we will put $k=0$ in the Einstein equations. More precisely, one can
show the inflationary scenario is satisfactory if the number of
e-folds, i.e. the logarithm of the scale factor at the end of
inflation to the scale factor at the beginning of inflation is greater
than $60$~\cite{inflation},
\begin{equation}
N_{_{\rm T}} >60 \, .
\end{equation}
More detailed arguments about the advantages of inflation can be found
in Ref.~\cite{procbrazil} but, at this point, it is important to
notice the following three facts. Firstly, inflation is convincing
because, by means of a single concept or hypothesis, one can solve
many different problems. In this sense, inflation is an economical
assumption. Secondly, as we will show below, inflation is falsifiable
since it makes definite predictions that we will describe. Therefore,
there is the hope either to confirm or to exclude this hypothesis and,
in any case, there is the certainty to learn something about the early
Universe. Thirdly, inflation is defined by the
condition~(\ref{definf}) but this does not prejudge the physical
nature of the matter responsible for the acceleration of the
Universe. The only thing one can say is obtained by expressing the
acceleration of the scale factor in cosmic time. Using the Einstein
equations, one gets
\begin{equation}
\label{accela}
\frac{\ddot{a}}{a}=-\frac{\kappa }{6}(\rho  +3p)\, ,
\end{equation}
where a dot denotes a derivative with respect to the cosmic
time. Therefore, the fluid responsible for inflation must be such that
\begin{equation}
p<-\frac{\rho }{3}\, ,
\end{equation}
i.e. must have a negative pressure. As a consequence, this fluid
cannot be a standard fluid, like a gas for instance, but must be
somehow ``exotic''. However, this does not come as a surprise since
inflation is supposed to take place at very high energies. At those
energies, the natural description of matter is (quantum) field
theory. As we are now going to demonstrate, it is quite interesting to
remark that the most simple example of a field theory can do the job
very well and ``produce'' the negative pressure which is necessary to
inflation. We now discuss this point in more details.

\subsection{Implementing the Inflationary Hypothesis}
\index{scalar field}

The most simple implementation of the inflationary scenario is to
assume that matter is described by a scalar field $\varphi (\eta
)$~\cite{guth,inflation}. This case is nothing but a particular
example of a perfect fluid. The corresponding action reads
\begin{equation}
S=-\int {\rm d}^4x\sqrt{-g} \biggl[\frac{1}{2}g^{\mu
\nu}\partial _{\mu} \varphi \partial _{\nu}\varphi +V(\varphi
)\biggr]\, .
\end{equation}
Then, the stress-energy tensor, which is defined by
\begin{equation}
T_{\mu \nu}=-\frac{2}{\sqrt{-g}}\frac{\delta S}{\delta g^{\mu \nu}}\, ,
\end{equation}
can be written as
\begin{equation}
T_{\mu \nu}=\partial _{\mu }\varphi \partial _{\nu }\varphi 
-g_{\mu \nu}\biggl[\frac{1}{2}g^{\alpha \beta } \partial _{\alpha
}\varphi \partial _{\beta }\varphi  +V(\varphi )\biggr]\, .
\end{equation}
From this expression, it is clear that the scalar field is indeed a
perfect fluid. The energy density and the pressure are defined by
$T^0{}_0=-\rho $, $T^i{}_j=p\delta ^i{}_j$ and we obtain
\begin{equation}
\label{rhopsf}
\rho = \frac{1}{2}\frac{(\varphi ')^2}{a^2}+V(\varphi ), \quad 
p=\frac{1}{2}\frac{(\varphi ')^2}{a^2}-V(\varphi )\, .
\end{equation}
The conservation equation can be obtained either by re-deriving it from
the very beginning or just by inserting the previous expressions of
the energy density and pressure into
Eq.~(\ref{conservation}). Assuming $\varphi '\neq 0$, this
reproduces the Klein-Gordon equation written in a FLRW background,
namely
\begin{equation}
\varphi ''+2\frac{a'}{a}\varphi '
+a^2\frac{{\rm d}V(\varphi )}{{\rm d}\varphi }=0\, .
\end{equation}
The other equation of conservation expresses the fact that the scalar
field is homogeneous and, therefore, does not bring any new
information. Finally, a comment is in order about the equation of
state.  In general, there is no simple link between $\rho $ and $p$
except when the kinetic energy dominates the potential energy, where
$\omega \equiv p/\rho \simeq 1$, i.e. the case of stiff matter or, on
the contrary, when the potential energy dominates the kinetic energy
for which one obtains $\omega \simeq -1$. This last case is of course
very interesting since this leads to an inflationary solution. We have
thus identified the condition under which inflation can occur: the
potential energy must dominate the kinetic energy, i.e.
\begin{equation}
V(\varphi )\gg \frac12 \frac{(\varphi ')^2}{a^2}\, .
\end{equation}
We now turn to a systematic study of this regime.

\subsection{Slow-roll Inflation}
\index{inflation!slow-roll}

Since the kinetic energy to potential energy ratio and the scalar
field acceleration to the scalar field velocity ratio are small, this
suggests to view these quantities as parameters in which a systematic
expansion is performed. The slow roll regime is controlled by the
three (at leading order) slow-roll parameters defined by:
\begin{eqnarray}
\label{defepsilon}
\epsilon &\equiv & 3 \frac{\dot{\varphi}^2}{2} 
\left(\frac{\dot{\varphi }^2}{2} + V\right)^{-1} = -\frac{\dot{H}}{H^2}
=1-\frac{{\cal H}'}{{\cal H}^2}\, ,
\\
\label{defdelta}
\delta &\equiv & -\frac{\ddot{\varphi }}{H\dot{\varphi }} =
- \frac{\dot{\epsilon }}{2 H \epsilon }+\epsilon\ , 
\quad 
\xi \equiv \frac{\dot{\epsilon }-\dot{\delta }}{H}\ .
\end{eqnarray}
The slow-roll conditions are satisfied if $\epsilon$ and $\delta$ are
much smaller than one and if $\xi = {\cal
O}(\epsilon^2,\delta^2,\epsilon\delta)$. It is also convenient to
re-express the slow-roll parameters in terms of the inflaton
potential. Using the equations of motion in the slow-roll
approximation, one can show that
\begin{equation}
\epsilon \simeq \frac{m_{_{\rm Pl}}^2}{16\pi}
\biggl(\frac{V'}{V}\biggr)^2, \quad 
\delta \simeq -\frac{m_{_{\rm Pl}}^2}{16\pi}
\biggl(\frac{V'}{V}\biggr)^2+\frac{m_{_{\rm Pl}}^2}{8\pi }\frac{V''}{V}\, ,
\end{equation}
where, here, a prime denotes a derivative with respect to the scalar
field. 

\par

The equations of motion, that is to say the Friedman equation and the
Klein-Gordon equation can be re-written exactly as
\begin{equation}
H^2=\frac{\kappa V}{3-\epsilon}\, ,\quad \frac{{\rm d}\varphi }{{\rm d}t}
=-\frac{1}{(3-\delta )H}\frac{{\rm d}V}{{\rm d}\varphi }\, ,
\end{equation}
from which one deduces that, if the slow-roll conditions are satisfied
\begin{equation}
\label{eqsr}
H^2\simeq \frac{\kappa }{3}V(\varphi )+{\cal O}\left(\epsilon
\right)\, ,\quad \frac{{\rm d}\varphi }{{\rm d}t} \simeq
-\frac{1}{3H} \frac{{\rm d}V}{{\rm d}\varphi }+{\cal O}\left(\delta
\right)\,.
\end{equation}
These equations are of course easier to analyze and solve than the
original ones.

\par

Let us now analyze a concrete example. We choose the following class
of potentials\index{slow-roll!polynomial potentials}
\begin{equation}
\label{pot}
V(\varphi )=\frac{3\lambda _n}{8\pi }m_{_{\rm Pl}}^4
\left(\frac{\varphi }{m_{_{\rm Pl}}}\right)^n\, ,
\end{equation}
where $n$ is a free parameter and $\lambda _n$ the coupling
constant. The factors that show up into the definition of the
potential have been chosen for future convenience. Let us first try to
see under which conditions the slow-roll approximation is valid. We
adopt the criterion $\epsilon <1$ (it is in fact $\epsilon \ll 1$ and
, strictly speaking, $\epsilon <1$ only corresponds to the condition
necessary in order to have an accelerated expansion). This amounts to
\begin{equation}
\varphi >\varphi _{\rm end}=\frac{n}{4\sqrt{\pi }}m_{_{\rm
Pl}}\, .
\end{equation}
Of course, this constraint applies in particular to the initial value
of the field. We already conclude, that for this class of models, the
values of the field must be at least a few Planck mass. Let us be more
precise and evaluate the total number of e-folds during slow-roll
inflation. It is given by the formula
\begin{equation}
\label{efold}
N_{_{\rm T}}=\ln \biggl(\frac{a_{\rm end}}{a_{\rm ini}}\biggr) \simeq
-\kappa \int _{\varphi _{\rm ini}} ^{\varphi _{\rm end}} {\rm
d}\varphi V(\varphi ) \biggl(\frac{{\rm d}V}{{\rm
d}\varphi}\biggr)^{-1}\, ,
\end{equation}
from which one gets
\begin{equation}
N_{_{\rm T}}=\frac{4\pi }{n}\left(\frac{\varphi _{\rm ini}}{m_{_{\rm
Pl}}}\right)^2 -\frac{n}{4}\, .
\end{equation}
Let $N_{\rm min}$ the minimum number of e-folds required in order to
solve the problems of the hot big-bang model (we have seen before that
$N_{\rm min}\simeq 60$) then one has
\begin{equation}
\varphi _{\rm ini}>m_{_{\rm Pl}}\sqrt{\frac{n}{4\pi }\left(N_{\rm min}
+\frac{n}{4}\right)}\, .
\end{equation}
For $n=2$, this gives $\varphi _{\rm ini}\simeq 3.1m_{_{\rm Pl}}$ and
for $n=4$, one obtains $\varphi _{\rm ini}\simeq 4.4m_{_{\rm
Pl}}$. However, it is often argued that ``natural'' initial conditions
(see the last article in Refs.~\cite{inflation}) are such that
$V(\varphi _{\rm ini})=m_{_{\rm Pl}}^4$ which amounts to
\begin{equation}
\label{ini}
\varphi _{\rm ini}=m_{_{\rm Pl}}\left(\frac{8\pi }{3}\right)^{1/n}
\lambda _n^{-1/n} \gg m_{_{\rm Pl}}\, , 
\end{equation}
because, as we will discuss later one, the COsmic Background Explorer
(COBE) normalization implies that the coupling constant is small. In
this case, the number of e-folds is a large number, much larger that
the minimum required. Of course, the fact that the value of the scalar
field must be larger or of the order of the Planck mass has led to
many discussions about the model building problem. In this review, we
do not address this question. Details about this issue can be found in
Refs.~\cite{LR,Linde}.

\par

Let us now solve the equations of motions in the slow-roll
approximation. For the scalar field straightforward calculations lead
to ($n\neq 4$)
\begin{equation}
\varphi (t)=\varphi _{\rm ini}
\left[1-\frac{n(4-n)}{2}\frac{\sqrt{\lambda _n}}{8\pi }
\left(\frac{m_{_{\rm Pl}}}{\varphi _{\rm ini}}\right)^{(4-n)/2}
m_{_{\rm Pl}}(t-t_{\rm ini})\right]^{2/(4-n)}\, .
\end{equation}
The last expression can also be expressed in terms of $t_{\rm end}$,
the time at which slow-roll inflation stops. One obtains
\begin{equation}
\varphi (t)=\varphi _{\rm ini}
\left\{1-\frac{t-t_{\rm ini}}{t_{\rm end}-t_{\rm ini}}
\left[1-\left(\frac{\varphi _{\rm end}}{\varphi _{\rm ini}}\right)
^{(4-n)/2}\right]\right\}^{2/(4-n)}\, .
\end{equation}
The advantage of the above equation is to show that for $t<t_{\rm
end}$ the argument between braces always remains positive and hence
the whole expression well-defined. A negative argument would simply
signal the break-down of the slow-roll approximation and, in this
case, the above formula cannot be used.

\par

Let us now turn to the scale factor. Integrating the first of
Eqs.~(\ref{eqsr}) leads to 
\begin{equation}
a(t)=a_{\rm ini}\exp \left\{-\frac{4\pi }{nm_{_{\rm Pl}}^2}
\left[\varphi _0^2(t)-\varphi _{\rm ini}^2\right]\right\}\, .
\end{equation}
From this expression, one can also calculate the evolution of the
scalar field in terms of the number of e-folds $N$ which is the
natural time variable during inflation. One gets
\begin{equation}
\varphi (N)=m_{_{\rm Pl}}\sqrt{\left(\frac{\varphi _{\rm ini}}{m_{_{\rm
Pl}}}\right)^2 -\frac{n}{4\pi }N}\, ,
\end{equation}
from which one obtains the formula giving the evolution of the 
Hubble parameter during inflation
\begin{equation}
\label{HN}
H(N)=m_{_{\rm Pl}}\sqrt{\lambda _n}\left[\left(\frac{\varphi _{\rm
ini}}{m_{_{\rm Pl}}}\right)^2 -\frac{n}{4\pi }N\right]^{n/4}\, .
\end{equation}
This equation is valid until $N=N_{_{\rm T}}$ and, in this regime, the
above formula is always well-defined. Indeed, Eq.~(\ref{HN}) becomes
meaningless at $N_{\rm max}=4\pi (\varphi _{\rm ini}/m_{_{\rm
Pl}})^2/n$ but $N_{\rm max}>N_{_{\rm T}}$. The above equation has
interesting consequences for our understanding of inflation. It shows
that the Hubble parameter can evolve and change significantly during
the slow-roll phase. Later on, we will see that a quantity which plays
an important role is the value of the Hubble parameter when the scales
of astrophysical interest today crossed out the horizon during
inflation. This happens $60$ e-folds before the end of inflation. This
scale is constrained by the observations on the Cosmic Microwave
Background Radiation (CMBR) anisotropies to be $H/m_{_{\rm
Pl}}<10^{-5}$. However, this does not mean that the Hubble parameter
has not been larger before, especially if the total number of total
e-folds is large, as it is the case for the initial conditions
discussed in Eq.~(\ref{ini}). For instance, if we have a massive
potential, $n=2$, and $\varphi _{\rm ini}=100\, m_{_{\rm Pl}}$ then
inflation starts with an initial Hubble parameter of $H_{\rm
ini}\simeq 10^{-3}m_{_{\rm Pl}}$ but ends at $H_{\rm end}=m_{_{\rm
Pl}} \sqrt{\lambda _n}\left[n/\left(4\sqrt{\pi
}\right)\right]^{n/2}\simeq 0.28\times 10^{-5}m_{_{\rm Pl}}$ after
$N\simeq 62000$ e-folds, where we have used $\lambda _2\simeq
10^{-10}$ (corresponding to a mass $m\simeq 10^{-5}m_{_{\rm
Pl}}$). Therefore, in summary, it will be important to keep in mind
that the observations give indications about the scale of inflation
when the relevant scales crossed out the horizon during inflation but
cannot, a priori, put constrains on the Hubble parameter in the
earliest phases of evolution.
\begin{figure}
\hspace{-1.2cm}
\includegraphics[angle=0,width=13cm]{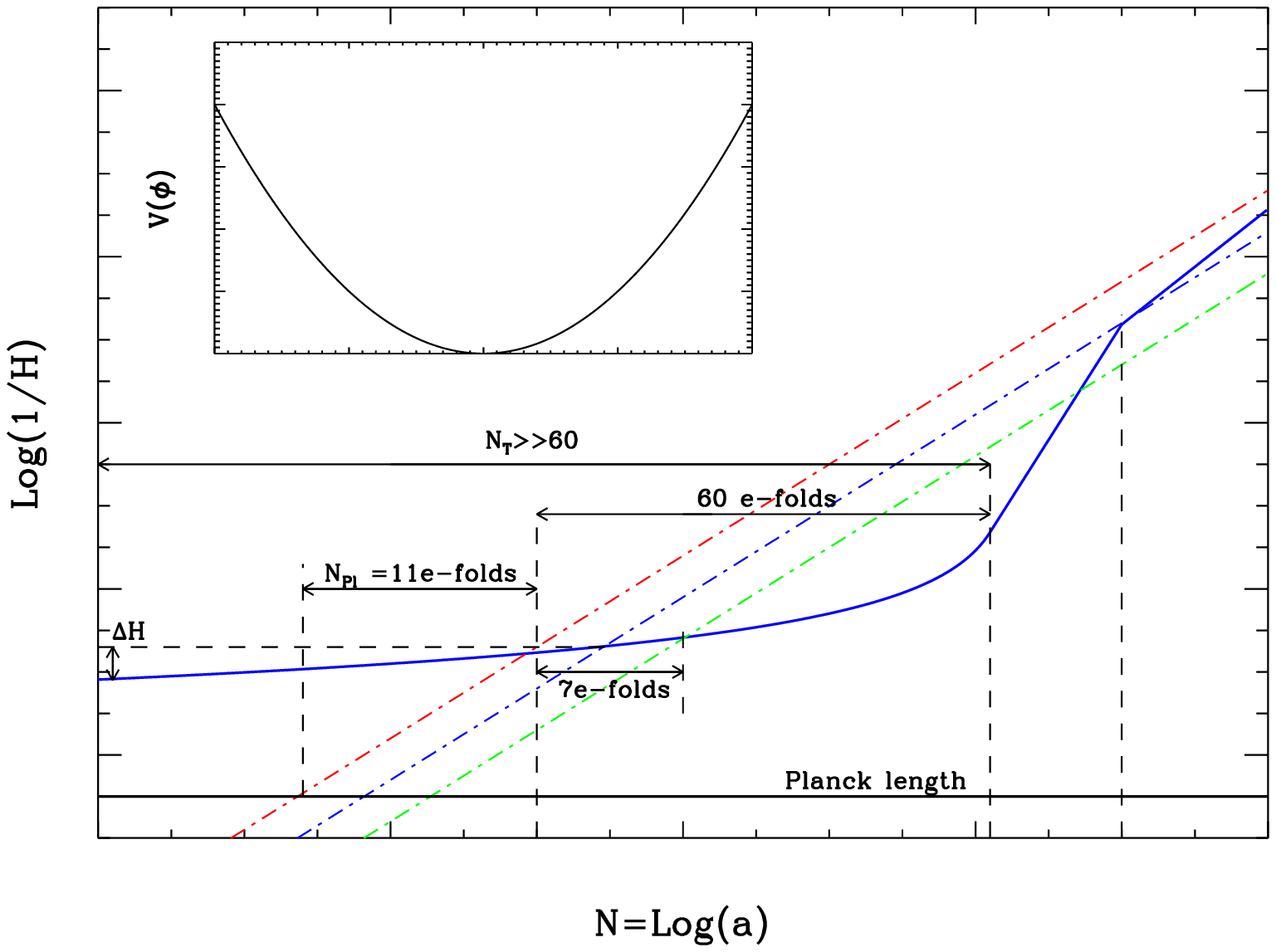}
\caption{Evolution of the various scales discussed in the text during
inflation and the subsequent radiation and matter dominated epochs. In
particular, it is apparent that the CMBR measurements only probe the
inflationary model when the modes of astrophysical interest today
crossed out the horizon during inflation. The small window shows a
typical inflationary potential, see Eq.~(\ref{pot}).}
\label{fig:scales}
\end{figure}
This situation is summarized in Fig.~\ref{fig:scales}.  

\par

When the field reaches the value $\varphi _{\rm end}$, slow-roll
inflation stops and the system enters a new regime that we now briefly
describe.

\subsection{Reheating}
\index{reheating!general theory}

When the scalar field reaches the point where the slow-roll parameter
$\epsilon \simeq 1$, for which $\varphi =\varphi _{\rm end}$,
inflation stops and the field starts oscillating around its
minimum~\cite{turner,preheating}. In this regime, the system is
governed by two time scales: the Hubble time $H^{-1}$ and the period
of the oscillations around the minimum $(V'')^{-1}$ (here, a prime
means derivative with respect to the field). The important point is
that these two scales are very different. The frequency of the
oscillations is much larger than the Hubble rate, $\omega _{\rm
osci}\simeq V'' \gg H$. The scalar field obeys the Klein-Gordon
equation which can be put under the following form
\begin{equation}
\frac{{\rm d}\rho }{{\rm d}t}
=-3H\dot{\varphi }^2=-6H(\rho -V)\, ,
\end{equation}
where the relation $\dot{\varphi}^2=2(\rho -V)$ has been used. This
equation can be time averaged and one gets
\begin{equation}
\label{avekg}
\biggl \langle \frac{{\rm d}\rho }{{\rm d}t} \biggr \rangle =
-\bigl \langle 6H(\rho -V) \bigr \rangle \simeq 
-6H \bigl \langle \rho -V \bigr \rangle \, ,
\end{equation}
where we have used the fact that the Hubble rate does not change
during one period of the oscillations. The right hand side of the
equation above can be evaluated as
\begin{equation}
\bigl \langle \rho -V \bigr \rangle 
\equiv \frac{1}{T}\int _0^T(\rho -V){\rm d}t
=\left[\int _{-\varphi _{\rm m}}^{\varphi _{\rm m}}
\sqrt{\rho -V(\varphi )}{\rm d}\varphi  \right] \left[
\int _{-\varphi _{\rm m}}^{\varphi _{\rm m}}
\frac{{\rm d}\varphi}{\sqrt{\rho -V(\varphi )}} \right]^{-1}\, ,
\end{equation} 
where $\varphi _{\rm m}$ is the value of the scalar field at the
maximum of its oscillations. In order to obtain the previous relation,
one has also utilized that ${\rm d}t={\rm d}\varphi / \sqrt{2(\rho
-V)}$. Then, one uses that over one period, $\rho \simeq V(\varphi
_{\rm m})\equiv V_{\rm m}$ is a constant and one obtains
\begin{equation}
\bigl \langle \rho -V \bigr \rangle \simeq \gamma \rho \, ,
\end{equation}
where the number $\gamma $ is defined by 
\begin{equation}
\gamma \equiv 
\left[\int _{-\varphi _{\rm m}}^{\varphi _{\rm m}}
\sqrt{1-\frac{V(\varphi )}{V_{\rm m}}}{\rm d}\varphi  \right] 
\left \{
\int _{-\varphi _{\rm m}}^{\varphi _{\rm m}}
\left[1-\frac{V(\varphi )}{V_{\rm m}}\right]^{-1/2}
{\rm d}\varphi \right \}^{-1} =\frac{n}{n+2}\, ,
\end{equation}
the last result being valid for potentials of the form $V(\varphi )
\propto \varphi ^{n}$. Let us now turn to the left hand side of
Eq.~(\ref{avekg}). The term $\langle {\rm d}\rho /{\rm d}t \rangle $
can be written as $\Delta \rho /T$. It can be expressed as a finite
difference expression and one can rewrite it as $\dot{\rho }$. This is
valid for time intervals much larger than the period of the
oscillations. Therefore, the equation governing the evolution of the
energy density of the field~(\ref{avekg}) can be rewritten as
\begin{equation}
\label{sol1}
\dot{\rho }=-\frac{6n}{n+2}H\rho \Rightarrow \rho \propto a^{-6n/(n+2)} \, .
\end{equation}
Then, the scale factor is given by $a(t)\propto t^{(n+2)/(3n)}$. For
the massive case, $n=2$, the energy density evolves as in a
matter-dominated epoch. This can be easily understood since, in this
case, the Klein-Gordon equation is exactly the equation of an harmonic
oscillator. In this situation, it is known that $\langle \dot{\varphi
}^2/2 \rangle =\langle V(\varphi )\rangle $ which implies that the
pressure vanishes.

\par

So far, we have not taken into account the effect of particles 
creation. Phenomenologically, it can be described by adding a term 
$\Gamma \dot{\varphi }$ in the Klein-Gordon equation which now reads 
\begin{equation}
\label{rhoGam}
\ddot{\varphi }+3H\dot{\varphi }+\Gamma \dot{\varphi }+
\frac{{\rm d}V(\varphi )}{{\rm
d}\varphi }=0 \quad \Rightarrow \frac{{\rm d}\rho }{{\rm d}t}
=-\frac{2n}{n+2}(3H+\Gamma )\rho .
\end{equation}
If we assume that the particles produced are very light in 
comparison with the mass of the inflaton, these particles will 
be very relativistic. This means that the equation of conservation 
of radiation should also be modified according to
\begin{equation}
\label{radGam}
\frac{{\rm d}\rho _{\rm r}}{{\rm d}t}
=-4H\rho _{\rm r} +\Gamma \rho \, ,
\end{equation}
so that the total energy is still conserved. Eq.~(\ref{rhoGam}) can be
easily integrated and the solution reads
\begin{equation}
\rho (t)=\rho _{\rm osci}\biggl(\frac{a}{a_{\rm osci}}\biggr)^{-6n/(n+2)}
\exp \biggl[-\frac{2n}{n+2}\Gamma (t-t_{\rm osci})\biggr]\, 
\end{equation}
where $t=t_{\rm osci}$ is the time at which the oscillations start
(i.e.  the time at which the slow-roll period ends) and $\rho _{\rm
osci}$ is the value of the scalar field energy density at that
time. The effect of the term $\Gamma \dot{\varphi }$ is to multiply
the result~(\ref{sol1}) by a decreasing exponential factor. Equipped
with this solution, we can solve Eq.~(\ref{radGam}) and determine the
evolution of $\rho _{\rm r}$. If the scalar field energy density
dominates the radiation, as it is the case at the beginning of the
reheating period, the solution reads (using the fact that the scale
factor is known in this regime, see above)
\begin{eqnarray}
\label{rhor1}
\rho _{\rm r}(t) 
&=& \Gamma t_{\rm osci}\rho _{\rm osci}
\biggl(\frac{a}{a_{\rm osci}}\biggr)^{-4}
\biggl(\frac{n+2}{2n\Gamma t_{\rm osci}}\biggr)^{(n+8)/(3n)}
\exp \biggl(\frac{2n}{n+2}\Gamma t_{\rm osci}\biggr)
\nonumber \\
& & \times 
\biggl[\gamma \biggl(\frac{n+8}{3n},\frac{2n}{n+2}\Gamma t\biggr)
-\gamma \biggl(\frac{n+8}{3n},\frac{2n}{n+2}\Gamma t_{\rm osci}\biggr)
\biggr]\, ,
\end{eqnarray}
where the function $\gamma (\alpha ,x)$ is the incomplete gamma
function defined by $\gamma (\alpha ,x)\equiv \int _0^x{\rm e}^{-t}
t^{\alpha -1}{\rm d}t$. In the above formula, we have assumed that, at
the end of the slow-roll epoch $t=t_{\rm osci}$, $\rho _{\rm r}\simeq
0$. For small values of the argument $x$, the incomplete gamma
function reduces to $\simeq x^{\alpha }/\alpha $. We define $t_{_{\rm
RH}}\equiv \Gamma ^{-1}$ and then we have $x=2n\Gamma
t/(n+2)=2n/(n+2)(t/t_{_{\rm RH}})$ and for times $t>t_{\rm osci} \ll
\Gamma ^{-1}$, the argument of the incomplete gamma function is
small. In this limit, one obtains
\begin{equation}
\label{rhoappro}
\rho _{\rm r}(t) \simeq 
\Gamma \rho _{\rm osci}t_{\rm osci }^2\frac{3n}{(n+8)t}
\biggl[1-\biggl(\frac{t}{t_{\rm osci}}\biggr)^{-(n+8)/(3n)}\biggr]\,.
\end{equation}
We see that $\rho _{\rm r}$ starts to increase, reaches a maximum and
then decreases. When $t$ approaches $t_{_{\rm RH}}$, the previous
approximation breaks down since the argument of the incomplete gamma
function is no longer small.  However, for order of magnitude
estimates, we can try to push this approximation. At $t\simeq t_{_{\rm
RH}}$, we have $\rho _{\rm r}\simeq \Gamma \rho _{\rm osci}t_{\rm
osci}^23n/[(n+8)t_{_{\rm RH}}]$ since the second term in the square
bracket in Eq.~(\ref{rhoappro}) is negligible. After thermalization,
the energy density of radiation takes the form $\rho _{\rm r}=g_*\pi
^2T^4/30$.  Using the fact that $\rho _{\rm osci}\simeq H_{\rm
inf}^2m_{_{\rm Pl}}^2$ (nothing but the Friedman equation) and that
$t_{\rm osci}\simeq H_{\rm inf}^{-1}$, one can deduce the reheating
temperature\index{reheating!temperature}
\begin{equation}
T_{_{\rm RH}}\simeq \frac{30^{1/4}}{\sqrt{\pi }} g_*^{-1/4}
\biggl(\frac{3n}{n+8}\biggr)^{1/4} (\Gamma m_{_{\rm Pl}})^{1/2}\, .
\end{equation}
Then, from this temperature, the universe evolves in a standard
radiation dominated era. The remarkable feature of the previous
equation is that it does not depend on the scale of inflation
$H_{_{\rm inf}}$ but only on the decay rate $\Gamma $ of the
inflaton. This means that whatever the scale of inflation is, the
radiation dominated era always starts at the same energy (at fixed
decay rate) and that the duration of the period of coherent
oscillations can change quite a lot. The number of e-foldings during
this epoch can be evaluated as [since during this epoch, the scale
factor scales as $\propto t^{(n+2)/(3n)}$]
\begin{equation}
\label{efoldosci}
N\simeq \frac{n+2}{3n}\ln \biggl(\frac{H_{\rm inf}}{\Gamma }\biggr) \, .
\end{equation}
The previous considerations are valid if the life time of the inflaton
is bigger than the age of the universe at the end of
inflation. Otherwise, there is no period of coherent oscillations. In
this case, the vacuum energy $H_{\rm inf}^2m_{_{\rm Pl}}^2$ is
directly converted into radiation and the reheating temperature is
\begin{equation}
T_{_{\rm RH}}\simeq \frac{30^{1/4}}{\sqrt{\pi }} g_*^{-1/4}(H_{\rm
inf}m_{_{\rm Pl}})^{1/2}\, .
\end{equation}
Finally, let us recall that the calculations above assume that the
physical quantities are time averaged and therefore that the time
scales considered are larger than the period of the oscillations. 
\begin{figure}
\hspace{-0.4cm}
\includegraphics[angle=0,width=12.5cm]{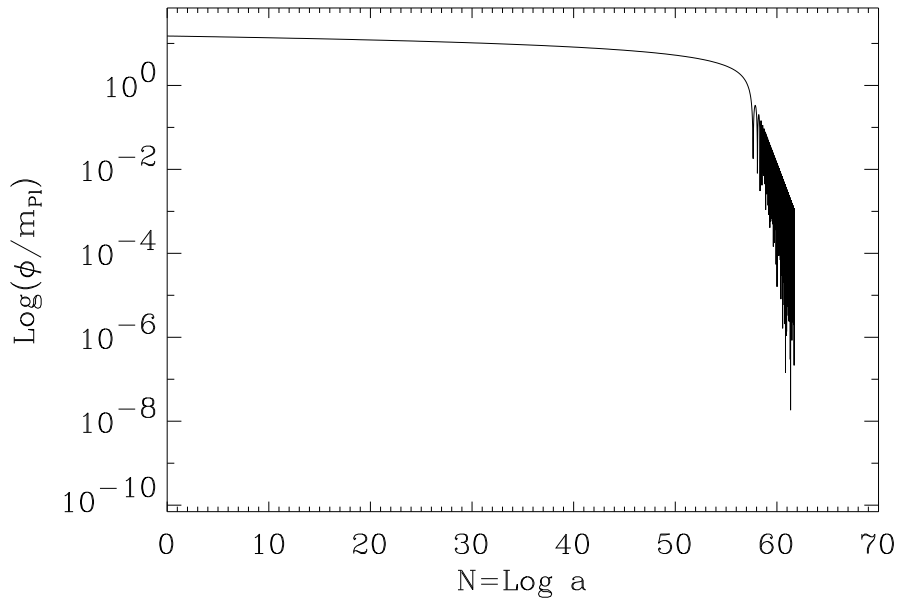}
\caption{Evolution of the scalar field during slow-roll inflation and
the reheating phase (where the field oscillates) obtained by numerical
integration of the equations of motion. The potential is of the type
of Eq.~(\ref{pot}) with $n=2$ and $\lambda _2=(8\pi /6)\times
10^{-10}$. The initial conditions are such that $\varphi _{\rm
ini}\simeq 3 m_{_{\rm Pl}}$ leading to $N_{_{\rm T}}\simeq 60$ as
confirmed by the plot.}
\label{fig:numerics}
\end{figure}
In Fig.~\ref{fig:numerics}, where the evolution of the field versus
the number of e-folds is displayed, we have integrated the equations
of motion numerically. This plot confirms our analytical estimates:
inflation consists of a phase of slow-roll followed by a phase of
oscillations. This concludes our study of the inflationary background.

\section{Cosmological Perturbations}
\label{sec:3}

\subsection{General Framework}
\index{cosmological perturbations!}

It is an observational fact that the universe is not isotropic and
homogeneous. Therefore, if one wants to have an accurate description,
it is clearly mandatory to go beyond the FLRW model. On the other
hand, it is also an experimental fact that, in the early Universe, the
deviations from the isotropy and from the homogeneity were small
(e.g. from the COBE measurement, $\delta T/T\simeq 10^{-5}$). This
suggests a perturbative treatment. Therefore, the following metric
tensor~\cite{MFB} gives a refined description of our Universe
\begin{equation}
\gamma _{\mu \nu }(\eta ,\vec{x})=[g_{\mu \nu}(\eta )+\epsilon h_{\mu
\nu }(\eta ,\vec{x}) +\epsilon ^2 \ell _{\mu \nu }(\eta
,\vec{x})+\cdots ]{\rm d}x^{\mu } {\rm d}x^{\nu }\, ,
\end{equation}
where $g_{\mu \nu}(\eta )$ is the standard FLRW metric introduced
previously and represents the ``background'' (the parameter $\epsilon
$ in the above equation should not be confused with the first
slow-roll parameter; they have nothing to do with each other). The
perturbed metric depends on $\vec{x}$ and this is the signature of the
fact that we now go beyond the cosmological principle. In order to be
consistent, the same expansion must be performed for the quantities
describing matter. For example, if there is a background scalar field
$\varphi (\eta )$, a refined description of the scalar field can be
expressed as
\begin{equation}
\varphi (\eta ,\vec{x})=\varphi (\eta )+\epsilon \delta \varphi
(\eta ,\vec{x}) +\epsilon ^2 \delta ^{(2)}\varphi (\eta ,\vec{x})
+\cdots \, .
\end{equation}
The main goal of the theory of cosmological perturbations is to
determine the evolution of the perturbed quantities $h_{\mu \nu}$ and
$\delta \varphi $ and, then, to use them in order to calculate
observable quantities. To find the behavior of the perturbed
quantities, one needs some equations of motion. Naturally, these
equations are taken to be the perturbed Einstein equations written
order by order (we assume that gravity is described by General
Relativity). Therefore, we expand the Einstein tensor and the
stress-energy tensor according to
\begin{equation}
G_{\mu \nu }=G_{\mu \nu }^{(0)}+\epsilon G_{\mu \nu }^{(1)}
+\epsilon ^2 G_{\mu \nu}^{(2)}+\cdots , \quad 
T_{\mu \nu}=T_{\mu \nu}^{(0)}+\epsilon T_{\mu \nu }^{(1)}
+\epsilon ^2T_{\mu \nu}^{(2)}+\cdots \, ,
\end{equation}
and then identify the terms of same order to obtain
\begin{equation}
G_{\mu \nu}^{(0)}=\kappa T_{\mu \nu}^{(0)}, \quad 
G_{\mu \nu}^{(1)}=\kappa T_{\mu \nu}^{(1)}, \quad 
G_{\mu \nu}^{(2)}=\kappa T_{\mu \nu}^{(2)}, \cdots \, . 
\end{equation}
In the present context, we will restrict ourselves to the linear order
in the parameter $\epsilon $.

\par

Let us now try to describe the perturbed metric tensor in more
details.  For any symmetric two-rank tensor, there is a
theorem~\cite{stewart} which states that $h_{\mu \nu}(\eta ,\vec{x})$
can be decomposed as $h_{\mu \nu}(\eta ,\vec{x})=h_{\mu \nu}^{\rm
(S)}+h_{\mu \nu}^{\rm (V)}+ h_{\mu \nu}^{\rm (T)}$, where $h_{\mu
\nu}^{\rm (S)}$ is constructed only from scalar functions, $h_{\mu
\nu}^{\rm (V)}$ is constructed only from three-dimensional vectors
with vanishing divergences and $h_{\mu \nu}^{\rm (T)}$ is obtained
only from transverse and traceless three-dimensional tensors. These
three types of perturbations are the scalar, rotational and tensorial
fluctuations respectively. Explicitly, the theorem implies that the
unperturbed metric plus the perturbed metric can be expressed
as~\cite{MFB}
\begin{eqnarray}
\label{ds2}
{\rm d}s^2 &=& a^2(\eta )\{- (1+2\phi){\rm d}\eta ^2 + 2(\partial
_iB-S_i) {\rm d}x^i {\rm d}\eta + [(1-2\psi)\delta _{ij}+2\partial _i
\partial _jE 
\nonumber \\ 
& & +\partial _jF_i + \partial
_iF_j+h_{ij}^{\rm (T)}]{\rm d}x^i{\rm d}x^j\} \ ,
\end{eqnarray}
with $S_i$ and $F_i$ being transverse vectors, i.e.~$\partial ^iS_i=
\partial ^iF_i= 0$ and $h_{ij}^{\rm (T)}$ being a transverse and
traceless tensor, i.e. $\delta ^{ij}h_{ij}=0$, $\partial ^jh_{ij}=0$.
We see that the scalar part of the metric depends on four unknown
functions: $\phi $, $B$, $\psi $ and $E$. The vector part depends on
two vectors with vanishing divergence, i.e. $S_i$ and $F_i$ and,
finally, the tensor part depends on one transverse and traceless
tensor, namely $h_{ij}^{\rm (T)}$. At linear order, each type of
perturbations decouple and, as a consequence, can be treated
separately.

\par

In the specific case of inflation, one can show that vector
perturbations cannot be produced~\cite{MFB}. Hence, in the following,
we will consider that only scalar and tensor perturbations are
present.

\par

At this point, one should discuss a well-known problem of the theory
of cosmological perturbations: the gauge issue. A complete study of
this question can be found in Refs.~\cite{MFB,Bardeen,MS1} but,
roughly speaking, it consists in the following. There exist solutions
to the perturbed Einstein equations which are coordinates dependent,
i.e.  which can be removed by performing an infinitesimal change of
coordinates. These solutions are fictitious and should not be
considered as physical. The following analogy may help to understand
the problem~\cite{RK}. Let us consider the four-dimensional FLRW
manifold denoted $V_4$ in what follows. It can be embedded into a
higher dimensional manifold, more precisely into the five-dimensional
Minkowski spacetime $E^5_{1,4}$ whose metric is $\eta _{AB}$ where the
indexes $A$ and $B$ runs from $0$ to $4$. A point in $E^5_{1,4}$ is
located by its coordinates $z^A$. An embedding is a map from $V_4$ to
$E^5_{1,4}$: $z^A=z^A(x^{\mu })$.  For a spatially flat FLRW spacetime
endowed with Cartesian coordinates, the embedding explicitly reads:
\begin{eqnarray}
z^0(\eta ,x,y,z) &=& \frac{1}{2}a(\eta )(x^2+y^2+z^2+1)+\frac{1}{2}
\int ^{\eta }\frac{a^2(\tau )}{a'(\tau )}{\rm d}\tau\, ,\\
z^1(\eta ,x,y,z) &=& \frac{1}{2}a(\eta )(x^2+y^2+z^2-1)+\frac{1}{2}
\int ^{\eta }\frac{a^2(\tau )}{a'(\tau )}{\rm d}\tau \, ,\\
z^2(\eta ,x,y,z) &=& ax, \quad z^3(\eta ,x,y,z)=ay, \quad 
\quad z^4(\eta ,x,y,z)=az\, .
\end{eqnarray}
Therefore, the FLRW manifold can be viewed as a surface into the
higher dimensional spacetime $E^5_{1,4}$. The metric of this surface
can be calculated by means of the well-known formula
\begin{equation}
\label{projection}
g_{\mu \nu}(\eta ,\vec{x})=\eta _{AB}\partial _{\mu }z^A\partial _{\nu
}z^B \, ,
\end{equation}
and we can indeed check that this reproduces the metric of a spatially
flat FLRW universe. Let us now try to ``deform'' this manifold since
this is what we have in mind when we consider small perturbations
around the background. In the present context, a deformation consists
of the following. If we consider a point $M$ in the manifold $V_4$
located by its coordinates $z^A(x^{\mu })$ in $E^5_{1,4}$, deforming
the manifold means slightly displacing the point $M$ in
$E^5_{1,4}$. This means that the coordinates of this point are no
longer $z^A$ but $z^A+\epsilon v^A(x^{\mu })$ where $\epsilon $ is a
small parameter. The vector $v^A(x^{\mu })$ characterizes the
deformation. As a consequence, the new metric of $V_4$ calculated by
means of Eq.~(\ref{projection}) reads
\begin{equation}
g_{\mu \nu }=\gamma _{\mu \nu}+2\epsilon \eta _{AB}\partial _{\mu }z^A
\partial _{\nu }v^B\, .
\end{equation}
However, all the vectors $v^A(x^{\mu })$ do not represent a 
genuine deformation. Indeed, if the following relation is satisfied 
\begin{equation}
z^A(x^{\mu })+\epsilon v^A(x^{\mu })
=z^A(x^{\mu }+\epsilon \xi ^{\mu })\, ,
\end{equation}
then, clearly, the displacement is within $V_4$ and does not
correspond to a deformation. This is merely a change of coordinates
that should not be considered as a physical deformation of $V_4$. This
gauge problem consists of identifying the spurious modes and in
removing them from the theory. To conclude this digression, it should
be emphasized that the link between the previous approach and the
theory of cosmological perturbations has never been worked out in
details. Therefore, an important warning is that it may well turn out
that the analogy used above cannot be applied completely to the theory
studied here.

\par

Having realized that there are non physical modes, the problem is now
to find a method to get rid of them. Following Bardeen's seminal
paper, an efficient way is to work with combinations of the metric
tensor components which are invariant under a general change of
coordinates (a ``gauge'' transformation) and, hence, which cannot
contain a spurious mode. For scalar perturbations, the two following
combinations~\cite{Bardeen}
\begin{equation}
\Phi (\eta ,\vec{x})\equiv \phi +\frac{1}{a}\biggl[a(B-E')\biggr]',\quad 
\Psi (\eta ,\vec{x})\equiv \psi  -\frac{a'}{a}(B-E')\, ,
\end{equation}
are gauge invariant. They are called the Bardeen potentials. In what
follows, we will see that, in the simple case where matter is
described by a scalar field, one has in fact $\Phi =\Psi$. This means
that we have reduced the study of the scalar perturbations to the
study of a single quantity: the Bardeen potential $\Phi (\eta
,\vec{x})$.

\par

The case of gravitational waves remains to be treated. In fact, it is
easy to realize that the gravitational waves are gauge-invariant by
definition because one cannot construct an infinitesimal change of
coordinates with a tensor. Therefore, one can safely work with the
tensor $h_{ij}^{\rm (T)}(\eta ,\vec{x})$ introduced before.

\par

We have identified the gauge invariant variables that describe the
gravitational sector. Our next move is to do the same but for the
matter sector. This can be done in general~\cite{Bardeen,MS1} but,
since we have inflation in mind, we just consider the case of a scalar
field. Then, one can show that
\begin{equation}
\delta \varphi ^{\rm (gi)}(\eta ,\vec{x})\equiv \delta \varphi +
\varphi '\left(B-E'\right)\, ,
\end{equation}
is the gauge-invariant perturbed scalar field. 

\par

Finally, we need a last ingredient. Since the spacelike sections are
flat and since we study the linear theory, it is very convenient to
work in the Fourier space. Indeed, because of the above properties,
each Fourier mode evolve independently (the mode coupling appearing at
quadratic order only) and it is sufficient to follow their time
evolution. Therefore, we Fourier transform the Bardeen potential and
the gravitational waves according to
\begin{eqnarray}
\Phi (\eta ,\vec{x}) &=& \frac{1}{(2\pi )^{3/2}}\int {\rm d}\vec{k} \Phi
(\eta ,\vec{k}){\rm e}^{i\vec{k}\cdot \vec{x}}\, ,
\\
h_{ij}^{\rm (T)}(\eta,\vec{x}) &=& \frac{1}{(2\pi )^{3/2}}
\int {\rm d}\vec{k}\sum _{s=+,\times }p_{ij}^s(\vec{k})
h_{_{\rm T}}^s(\eta ,\vec{k}){\rm e}^{i\vec{k}\cdot \vec{x}}\, .
\end{eqnarray}
In the last equation, $p_{ij}(\vec{k})$ is the transverse and
traceless polarization tensor satisfying the following properties:
$p_{ij}^s(\vec{k})p^{ij}{}^{s'}(\vec{k})=2\delta ^{ss'}$. The symbols
``$+$'' and ``$\times$ '' denote the two possible states of
polarization of the gravitational waves. Of course, we also Fourier
transform the perturbed scalar field and work with $\delta \varphi
^{\rm (gi)}(\eta ,\vec{k})$.

\par

Having identified what the relevant degrees of freedom are, we now
turn to the question of establishing their equation of motion. 

\subsection{Equations of Motion}

Since the Einstein equations are obviously gauge-invariant ``by
definition'', it is clear that it is possible to express them in terms
of gauge invariant quantities only. We start with density
perturbations.  Lengthy but straightforward calculations lead to (for
a fixed Fourier mode $k$)
\begin{eqnarray}
\label{2-11}
& & -3{\cal H}({\cal H}\Phi +\Psi') - k^2 \Psi =
\frac{\kappa }{2}\left\{-(\varphi ')^2 \Phi + 
\varphi '\left[\delta \varphi ^{\rm (gi)}\right]'+
a^2 \frac{{\rm d}V}{{\rm d}\varphi }
\delta \varphi ^{\rm (gi)}\right \}\, , 
\\
& & {\cal H}\Phi +\Psi' = \frac{\kappa }{2} \varphi '
\delta \varphi ^{\rm (gi)} \, , \quad
\Phi - \Psi = 0 \, , 
\\
& & (2{\cal H}'+{\cal H}^2)\Phi +
{\cal H}\Phi '+\Psi ''+2{\cal H}\Psi'-
\frac{1}{3} k^2 (\Phi -\Psi) = 
\frac{\kappa }{2}\biggl \{-(\varphi ')^2 \Phi 
\nonumber \\
& & +
\varphi '\left[\delta \varphi ^{\rm (gi)}\right]'
-a^2 \frac{{\rm d}V}{{\rm d}\varphi }
\delta \varphi ^{\rm (gi)}\biggr \}\, .
\end{eqnarray}
As announced, the two Bardeen potentials $\Phi $ and $\Psi $ are
equal. This is true as long as there is no anisotropic stress. Despite
the apparent complexity of this system of equations, straightforward
manipulations show that everything can be reduced to the study of a
single equation which reads
\begin{equation}
\label{bardeenmaster}
\Phi'' + 2\biggl({\cal H}-\frac{\varphi ''}{\varphi '}\biggr)\Phi' +
\biggl[k^2 + 2\biggl({\cal H}'-{\cal H}\frac{\varphi ''} {\varphi
'}\biggr)\biggr]\Phi = 0 \, .
\end{equation}
This equation is valid provided $\varphi '\neq 0$. In this case, for
which the scalar field plays the role of a cosmological constant, we
have no density perturbations at all, $\Phi =0$. This does not mean
that the perturbed scalar field cannot fluctuate in de Sitter
spacetime (as a matter of fact, it does) but that, in this case, these
fluctuations do not couple to the fluctuations of the
metric. Eq.~(\ref{bardeenmaster}) can be transformed in order to
permit a more transparent physical interpretation. If we consider the
variables $u$ and $\theta $ defined by
\begin{equation} 
\label{sigma} 
u\equiv\frac{4}{3}{a^2\theta\over{\cal H}}\Phi \ ,
\qquad \theta \equiv \frac1a \left(\rho \over \rho + p\right)^{1/2}
= \sqrt{3}{{\cal H}\over a\varphi '} 
=\sqrt{\frac{3}{2}}\frac{1}{a\sqrt{\gamma }}\, ,
\end{equation}
then Eq.~(\ref{bardeenmaster}) can be expressed as:
\begin{equation}
\label{equ}
u''+\biggl(k^2-\frac{\theta ''}{\theta }\biggr)u=0 \, .
\end{equation}
The above equation can be viewed either as the equation of a
parametric oscillator, with a time-dependent frequency given by
$\omega ^2(\vec{k},\eta )\equiv k^2-\theta ''/\theta $, or as a
Schr\"odinger equation with the potential $\theta ''/\theta $. This
effective potential contains derivatives of the scale factor up to the
fourth order. The typical behavior of the solutions can be easily
found. For modes $k^2\gg \theta ''/\theta $, the variable $u$
oscillate, $u\propto {\rm e}^{ik\eta }$, while for modes $k^2 \ll
\theta''/\theta$ the solution of Eq.~(\ref{equ}) may be expanded in
powers of $k^2$. At leading order we obtain
\begin{equation}
\label{sigmaexp}
u(\eta ,\vec{k})=A_1(k) 
\theta (\eta )\int ^{\eta }{{\rm d}\tau \over \theta^2(\tau)}+ 
A_2(k)\theta (\eta ) \, .
\end{equation}
Since $\theta \to \infty$ for $a \to 0$ in general, $A_1$ is the
arbitrary constant in front of the regular (growing) mode and $A_2$ a
constant associated with the singular (decaying) mode. We will see in
the following that the variable $u$ is in fact not the most
interesting for density perturbations. Quantum-mechanical
considerations, among others, will lead us to work with another
variable. From the above solution for $u(\eta ,\vec{k})$, we easily
deduce the Bardeen potential in the superhorizon regime. One obtains
\begin{equation}
\label{sigmagrm}
\Phi (\eta ,\vec{k})\simeq -A_1(k) {{\cal H}\over 2a^2} 
\int^\eta a^2 \gamma (\tau ){\rm d}\tau  +\frac{A_2(k)}{2k^2}
\frac{{\cal H}}{a^2}\, .
\end{equation}
For example, for a power-law behavior of the scale factor, i.e.  $a
\propto |\eta|^{1+\beta}$ with $\beta \le -2$ in order to have
inflation, the `growing' mode turns out to be constant in time, namely
\begin{equation}
\Phi (\eta ,\vec{k})\simeq \frac{3}{2} 
{1+\omega \over 5+3\omega }A_1(k) \, . 
\end{equation}
For the de Sitter case, $\omega =-1 $ and we recover the fact that
there no density perturbations at all.

\par

The equation of motion (\ref{bardeenmaster}) has a first integral for
modes that are much larger than the Hubble scale, i.e.  $k/a \ll
H$. Following Ref.~\cite{MFB} we define
\begin{equation}
\label{defzeta}
\zeta \equiv \frac 23 {{\cal H}^{-1} \Phi^\prime 
+ \Phi\over 1 + \omega } + \Phi \ ,
\end{equation}
which was introduced by Lyth \cite{L} originally. Essentially, the
quantity $\zeta$ is the perturbation of the intrinsic curvature in the
comoving gauge \cite{L} and is written $-{\cal R}$ in that
reference. The equation of motion for the Bardeen potential can be
re-written as an expression for the first derivative of the quantity
$\zeta $. One obtains~\cite{MS1}
\begin{equation}
\label{zp}
{1\over {\cal H}} \frac{{\rm d}\zeta }{{\rm d}\eta }\propto 
\left(k\over {\cal H}\right)^2 \Phi \, .
\end{equation}
Of course, to derive this equation we have not assumed that the
equation of state parameter is a constant. Thus, $\zeta$ is constant
in time for superhorizon modes $k/{\cal H} \ll 1$ since then
$\dot{\zeta}=0$.

\par

The importance of the quantity $\zeta $ is due to the fact that this
is a pure geometrical quantity. Concretely, this means that the
conservation law established above in the case of a scalar field is in
fact valid for any type of matter (at least, provided that the
so-called entropy perturbations do not play an important
role). Therefore, $\zeta $ can be used as a ``tracer'' of density
perturbations regardless of the type of matter responsible for those
fluctuations. In particular, it can be used to propagate the spectrum
from the end of inflation (where the Universe is dominated by a scalar
field) to the radiation dominated era (where the Universe is dominated
by a relativistic fluid) without knowing the details of the reheating
process. Let us now study how the calculation works in details. On
superhorizon scales, the Bardeen potential is almost constant. If we
neglect its time derivation then the equation of motion for $\Phi $
and the definition of $\zeta $ lead to
\begin{equation}
\Phi \simeq \frac{3(1+\omega _{_{\rm inf}})}{5+3\omega _{_{\rm
inf}}}\zeta \simeq \frac{\kappa \varphi '}{2{\cal H}}\delta \varphi
^{\rm (gi)}\, .
\end{equation}
Using the equation of motion of the background, the above formula 
can be put under the following form
\begin{equation}
\label{spectrumzeta}
\zeta _{_{\rm inf}}(\eta ,\vec{k})=\frac{5+3\omega _{_{\rm
inf}}}{2}{\cal H} \left[\frac{\delta \varphi ^{\rm (gi)}(\eta
,\vec{k})}{\varphi '} \right]\, .
\end{equation}
Now, let us assume that we want to know the Bardeen potential in an
era dominated by a fluid with a given equation of state $\omega $
(concretely, we have in mind $\omega =1/3$ or $\omega =0$ for the
radiation or matter dominated epochs, respectively). Writing the
constancy of $\zeta $, i.e. $\zeta _{_{\rm inf}}=\zeta _{\omega }$,
one arrives at
\begin{equation}
\Phi _{\omega }(\eta ,\vec{k})\simeq 3{\cal H}
\left(\frac{1+\omega }{5+3\omega }\right)
\left[\frac{\delta \varphi ^{\rm (gi)}(\eta
,\vec{k})}{\varphi '} \right]\, ,
\end{equation}
where we have used $\omega _{_{\rm inf}}\simeq -1$. This equation is
very important since it links the primordial fluctuations of the
scalar field to the fluctuations of the gravitational potential during
the subsequent phases of evolution of the Universe. 

\par

Let us now turn to gravitational waves. In order to obtain the
equation of motion, we must compute the perturbed Ricci or Einstein
tensors for the metric given in Eq.~(\ref{ds2}). One finds that
$\delta R^{0}{}_0=\delta R^{0}{}_i=0$. This result is consistent with
the fact that the gravitational waves are transverse and traceless
since only the trace and/or the derivative of the metric tensor can
appear in these components. On the other hand, the component $\delta
R^{i}{}_j$, $i\neq j$ is non vanishing and the leads to
\begin{eqnarray}
\delta R^{i}{}_j &=& \frac{1}{2a^2}\left[h^{\rm (T)}{}^i{}_j\right]''
+\frac{a'}{a^3}\left[h^{\rm (T)}{}^i{}_j\right ]'
-\frac{1}{2a^2}\partial _k\partial
^kh^{\rm (T)}{}^i{}_j=\kappa \delta T^i{}_j\, ,
\end{eqnarray}
where $\delta T^i{}_j$ represents the anisotropic pressure part of the
perturbed stress-energy tensor. For a perfect fluid (e.g. for a scalar
field), it vanishes. However, it is important to keep in mind that, a
priori, the gravitational waves are not ``independent'' from the
matter fluctuations and, hence, that they should be considered on the
same footing as density perturbations. This is conceptually important
because this means that it would be incorrect to argue that both types
of perturbations should be treated differently, in particular with
respect to the quantization of the cosmological perturbations.

\par

Then, for the rescaled Fourier amplitude defined by $\mu _{_{\rm
T}}^s(\eta ,\vec{k})\equiv a(\eta )h^s(\eta ,\vec{k})$, the equation
$\delta R^i{}_j=0$ can be re-written as~\cite{Gri}
\begin{equation}
\label{eomgw}
\left(\mu _{_{\rm T}}^s\right)''+\biggl(k^2-\frac{a''}{a}\biggr)\mu
_{_{\rm T}}^s= 0\, .
\end{equation}
This equation can be viewed either as the equation of a parametric
oscillator, i.e. an oscillator with a time-dependent frequency,
$\omega ^2(\eta ,\vec{k})\equiv k^2-a''/a$ or as a
``time-independent'' Schr\"odinger equation with a potential $U_{_{\rm
T}}(\eta )=a''/a$.  Therefore, we obtain the same type of equation as
for density perturbations. However, it is also interesting to notice
that the effective potential for tensor perturbations involves the
scale factor and its derivatives up to second order only. This
difference is especially important during the reheating phase.

\par

As it was the case for density perturbations, it is clear that the
solution to the equation of motion possesses two regimes. If the wave
number $k$ is such that $k^2\gg U_{_{\rm T}}$, then the mode function
oscillates, i.e.  $\mu _{_{\rm T}}\propto {\rm e}^{ik\eta }$. The
interaction with the barrier (if any) corresponds to the time $k\eta
_{\rm t}\simeq 1$ since, in the inflationary phase, one has $U_{_{\rm
T}}\simeq 1/\eta ^2$ (for a power-law scale factor; this is also the
case for slow-roll inflation, see below). This time is also, roughly
speaking, the time of Hubble radius exit. Indeed, the wavelength is
given by $\lambda =2\pi a(\eta )/k$ and the Hubble scale is
$H^{-1}=a/{\cal H}$. The condition $\lambda =H^{-1}$ gives $k\eta
_{\rm t}\simeq 1$ since ${\cal H}\simeq 1/\eta $. The second regime is
when the wave is below the potential, $k^2\ll U_{_{\rm T}}$. An
approximate solution is
\begin{equation}
\label{supmu}
\mu _{_{\rm T}}^s(\eta ,\vec{k})\simeq B_1^s(k)a(\eta )+B_2^s(k)a(\eta
)\int ^{\eta } \frac{{\rm d}\tau }{a^2(\tau )}\, ,
\end{equation}
where $B_1^s(k)$ and $B_2^s(k)$ are two constants which are {\it a
priori} free. The first term in Eq.~(\ref{supmu}) is the growing mode
whereas the second term is the decaying mode. This can be seen, for
example, if we consider scale factors of the form $a(\eta )=\ell
_0\vert \eta \vert ^{1+\beta }$. In this case, $\mu _{_{\rm T}}^s
\simeq B_1^s(k)\vert \eta \vert ^{1+\beta }+B_2^s(k)\vert \eta \vert
^{-\beta }$ and for $\beta \simeq -2$, the first term goes to infinity
while the conformal time goes to zero at the end of inflation.
Therefore, this is indeed the growing mode. In terms of the amplitude
$h^s(\eta ,\vec{k})$ itself, one sees that the growing mode
corresponds in fact to a constant and hence is conserved on large
scales. Somehow, $h^s(\eta ,\vec{k})$ plays for gravitational waves
the same role as $\zeta $ for density perturbations. 

\par 

Let us end this section with a comparison between density
perturbations and gravitational waves. Using the results obtained
before, the tensor to scalar amplitudes ratio today is given by
\begin{equation}
\frac{h_{\omega }}{\Phi _{\omega }}\sim (1+\omega _{_{\rm inf}})
\frac{h_{_{\rm inf}}}{\Phi _{_{\rm inf}}}\, .
\end{equation}
Therefore, if we assume that $h_{_{\rm inf}}\simeq \Phi _{_{\rm
inf}}$, which is the case if the perturbations are of
quantum-mechanical origin, then we have $h_{\omega }/\Phi _{\omega
}\ll 1$, i.e. scalar fluctuations dominates over tensor fluctuations,
because during inflation $\omega _{_{\rm inf}}\simeq -1$.

\par

The previous considerations also illustrate the limitations of the
classical approach. Without a theory of the initial conditions, i.e.
without a prescription to choose the $k$-dependent constants
$A_{1,2}(k)$ for density perturbations and $B_{1,2}(k)$ for
gravitational waves, we cannot really go further. This will be one of
the main advantage of the quantum-mechanical version of the previous
theory: a natural choice for $A_{1,2}(k)$ and $B_{1,2}(k)$.

\subsection{The Sachs-Wolfe Effect}
\index{Sachs-Wolfe effect}

The production and the amplification of small inhomogeneities in the
early Universe described above has several observational
consequences. In this review, we focus on one of them: the presence of
small angular anisotropies in the temperature of the Cosmic Microwave
Background Radiation (CMBR), at the level of $\delta T/T\simeq
10^{-5}$, detected for the first time by the COBE satellite in
$1992$~\cite{cobe}. These anisotropies are of utmost importance for
the theory of inflation because they allow us to check the predictions
of this scenario and/or to constrain the physics of the early
Universe. We now turn to a rapid discussion of this effect, a complete
presentation being available in Ref.~\cite{pdb}.

\par

The Sachs-Wolfe effect~\cite{sw} links the angular variations of the
temperature on the celestial sphere to the presence of cosmological
fluctuations in the early Universe. We have to calculate the change in
the energy of the photons propagating from the last scattering surface
to Earth.  This energy is given by $E=-\gamma _{\mu \nu }u^{\mu
}k^{\nu }$, where $k^{\mu }$ is the wave vector of the photon and
$u^{\mu }$ the velocity of the observer. Let us first investigate this
relation for the background.

\par

Fundamental observers are observers who move with the cosmological
flow.  A trajectory is given by the set $x^{\mu }=x^{\mu }(s)$, where
$s$ is a affine parameter along the line. The velocity along this
curve is given by $u^{\mu }\equiv {\rm d}x^{\mu }/{\rm d}s$ and
satisfies $u^{\mu }u_{\mu }=-1$. For a fundamental observer, one has
$u^i=0$ by definition and the normalization of the four velocity
implies that $u^{\mu }=(1/a,0)$ and $u_{\mu }=(-a,0)$. Let us now
study the propagation of a photon.  If $k^{\mu }\equiv {\rm d}x^{\mu
}/{\rm d}\lambda $ is the wave vector of a photon, then the path
followed by this photon is such that
\begin{equation}
\label{eqphoton}
\frac{{\rm d}k^{\mu }}{{\rm d}\lambda }+\Gamma ^{\mu }_{\nu \rho}k^{\nu }
k^{\rho }=0\, .
\end{equation}
The solution of this equation is the trajectory of the photon: $x^{\mu
} =x^{\mu }(\lambda )$. In addition, we have the constrain $k^{\mu
}k_{\mu }=0$, expressing the fact that the photon follows a null
geodesic. At zeroth order, this constraint gives $\delta
_{ij}k^{i}k^{j}=\left(k^{0}\right)^2$. In the non perturbed universe,
Eq.~(\ref{eqphoton}) possesses the solutions $k^{0}=C^0/a^2$, where
$C^0$ is a constant and $k^{i}=-C^0e^i/a^2$, where $e^i$ is a three
vector such that ${\rm d}e^i/{\rm d}\lambda =0$. Taking the ratio of
the wave vector components, we deduce that ${\rm d}x^i/{\rm d}\eta
=-e^i$. Finally, integrating this relation, we find the equation of
the trajectory in the unperturbed Universe
\begin{equation}
x^i=-e^i(\eta -\eta _{_{\rm D}})+x^i_{_{\rm D}}\, ,
\end{equation}
where $(\eta _{_{\rm D}},x^i_{_{\rm D}})$ are the coordinates at
detection of the photon. It does not come as a surprise that the
photons propagate along a straight line. On the other hand, the energy
is given by
\begin{equation}
E(\eta )=\frac{C^0}{a(\eta )}\, ,
\end{equation}
and, in fact, we just recover the well-known time evolution of the
temperature (which, as expected, does not depend on space for the
unperturbed Universe).

\par

Let us now turn to the Sachs-Wolfe effect itself. Essentially, this
consists in computing the energy of the photons at first order. In a
perturbed Universe, the most general observer possesses a velocity
given by $u^{\mu }+\delta u^{\mu }$, where $u^{\mu }$ denotes the
velocity of a fundamental observer calculated above and where we
assume that the components of $\delta u^{\mu }$ are small with respect
to this fundamental velocity. The fact that the total velocity is
normalized to $-1$ implies that $\delta u^{0}=-\phi /a$. We also write
$\delta u^{i}$ as $\delta u^{i}\equiv v^i/a$, from which we deduce
that $\delta u _i=av_i+a\partial _iB$. The trajectory of the photons
can also be expanded as $x^{\mu }+\delta x^{\mu }$, where $x^{\mu }$
is the path of the photon in an unperturbed Universe determined before
and $\delta x^{\mu }$ are the small corrections around the background
trajectory due to the presence of the fluctuations. In the same manner
as we did for the four-velocity, we can expand the wave vector of the
photon according to $k^{\mu }+\delta k^{\mu }$, where $\delta k^{\mu
}\equiv {\rm d}\left(\delta x^{\mu }\right)/{\rm d}\lambda $. At first
order, the variation of energy can be expressed as
\begin{equation}
\delta E =-h_{\mu \nu}u^{\mu }k^{\nu }-g_{\mu \nu} \delta u^{\mu
}k^{\nu }-g_{\mu \nu} u^{\mu }\delta k^{\nu }\, ,
\end{equation}
which can be re-written as $\delta E=C^0\phi/a+C^0e^i(\partial
_iB+v_i)/a +a\delta k^{0}$. In this equation, the only unknown
quantity is $\delta k^0$ and we now establish its
expression. Integrating the perturbed version of Eq.~(\ref{eqphoton}),
one finds that
\begin{eqnarray}
\delta k^{0} &=& -\frac{C^0}{a^2(\eta )}
\int _{\eta _{_{\rm E}}}^{\eta }
{\rm d}\tau \biggl\{\phi '-2e^i\partial _i\phi-e^ie^j
\partial _i\partial _jB
+\frac{1}{2}\biggl[-2\psi\delta _{ij}+2\partial _i\partial _jE
\nonumber \\
& & +h_{ij}^{\rm (T)}\biggr] '\, e^ie^j\biggr\}\, ,
\end{eqnarray}
where $\eta _{_{\rm E}}$ is the conformal time at emission. Let us
stress again that the integration is performed along the unperturbed
path of the photon. Putting everything together, we finally obtain
\begin{eqnarray}
\label{ratioE}
\frac{E_{_{\rm D}}}{E_{_{\rm E}}} &=& \frac{a_{_{\rm E}}}{a_{_{\rm D}}}
\biggl\{ 1+\left[\phi +
e^i\left(\partial _iB+v_i\right)\right]_{_{\rm E}}^{_{\rm D}}
-
\int _{\eta _{_{\rm E}}}^{\eta _{_{\rm D}}}
{\rm d}\tau \biggl[\phi '-2e^i\partial _i\phi-e^ie^j
\partial _i\partial _jB
\nonumber \\
& & +\frac12\left(-2\psi\delta _{ij}+2\partial _i\partial _jE
+h_{ij}^{\rm (T)}\right)'e^ie^j\biggr]\biggr\} \, .
\end{eqnarray}
The above expression depends on the coordinates of emission and
detection of the photons. To go further, it is necessary to specify
the conditions of emission, that is to say the characteristics of the
last scattering surface. At zeroth order, the surface of last
scattering has coordinates $\eta _{_{\rm E}}=\eta _{\rm lss}$,
$x^i_{_{\rm E}}=-e^i(\eta _{\rm lss}-\eta _{_{\rm D}})+x^i_{_{\rm D}}$
where $\eta _{\rm lss}$ is fixed and corresponds to the redshift
$z_{\rm lss}\simeq 1100$. The only dependence is now the vector $e^i$
and this corresponds to different directions on the celestial
sphere. However, in presence of perturbations, emission occurs at
different times and at different positions. In other words, the time
of emission is given by $\eta _{_{\rm E}}=\eta _{\rm lss}+\delta \eta
(\eta _{\rm lss},x^i_{_{\rm E}})$. The quantity $\delta \eta (\eta
_{\rm lss},x^i_{_{\rm E}})$ depends on our definition of the surface
of emission. Let us assume that this surface is such that the density
of photons, $\rho _{\gamma }$, is constant. Writing this condition at
first order gives $\delta \rho _{\gamma }(\eta _{\rm lss},x^i_{_{\rm
E}}) +\rho _{\gamma }'(\eta _{\rm lss})\delta \eta (\eta _{\rm
lss},x^i_{_{\rm E}})=0$. Using the conservation equation which implies
that $\rho _{\gamma }'=-4{\cal H}\rho _{\gamma }$, we arrive at
\begin{equation}
\delta \eta (\eta _{\rm lss},x^i_{_{\rm E}})=
\frac{1}{4{\cal H}(\eta _{\rm lss})}
\frac{\delta \rho _{\gamma }(\eta _{\rm lss},x^i_{_{\rm E}})}{
\rho _{\gamma }(\eta _{\rm lss})}\, .
\end{equation}
Therefore, the term $a(\eta _{_{\rm E}})$ in Eq.~(\ref{ratioE}) should
be written as
\begin{equation}
a(\eta _{_{\rm E}})=a(\eta _{\rm lss})+{\cal H}(\eta _{\rm lss})
\delta \eta (\eta _{\rm lss},x^i_{_{\rm E}})=a(\eta _{\rm lss})
+\frac{1}{4}\delta _{\gamma }(\eta _{\rm lss},x^i_{_{\rm E}})\, ,
\end{equation}
where $\delta _{\gamma }\equiv \delta \rho _{\gamma }/\rho _{\gamma }$
is the density contrast. In the same manner, if we say that detection
takes place on a surface such that the baryons energy density is
constant, the factor $a^{-1}(\eta _{_{\rm R}})$ should be written as
$a^{-1}(\eta _{_{\rm D}}) =a^{-1}(\eta _0)[1-(1/3)\delta _{\rm b}(\eta
_0,x^i_{_{\rm D}})]$ (the factor $1/3$ comes from the equation of
conservation but now written for a fluid whose equation of state
vanishes). Finally, Eq.~(\ref{ratioE}) takes the form
\begin{eqnarray}
\label{ratioEfinal}
\frac{E_{_{\rm D}}}{E_{_{\rm E}}} &=&
\frac{a(\eta _{\rm lss})}{a(\eta _0)}
\biggl\{ 1+\frac{1}{4}\delta _{\gamma }(\eta _{\rm lss},x^i_{_{\rm E}})
-\frac{1}{3}\delta _{\rm b}(\eta _0,x^i_{_{\rm D}})+
\left[\phi +e^i(\partial _iB+v_i)\right]_{_{\rm E}}^{_{\rm D}}
\nonumber \\
& & - \int _{\eta _{\rm lss}}^{\eta _{0}}
{\rm d}\tau \biggl[\phi '-2e^i\partial _i\phi-e^ie^j
\partial _i\partial _jB
+\biggl(-\psi\delta _{ij}+\partial _i\partial _jE
\nonumber \\
& & +\frac12h_{ij}^{\rm (T)}\biggr)'e^ie^j\biggr]\biggr\} \, .
\end{eqnarray}
Having established this important relation, we must now show that this
expression is gauge-invariant. For this purpose, it is sufficient to
express the ratio of the energies at emission and detection only in
terms of gauge-invariant quantities. We have already described the
gauge-invariant variables for the gravity sector. For the variables
describing matter, we only need the gauge-invariant density contrast
$\delta _{\rm g}\equiv \delta +\rho '/\rho (B-E')$. Finally, one has
to decompose the three-velocity as $v_i=\partial _iv$ and the
gauge-invariant velocity can be expressed as $v^{\rm (gi)}\equiv
v+E'$. Let us also notice that the spatial derivatives can be
expressed in terms of time derivatives. Indeed, along a trajectory,
one has ${\rm d}f/{\rm d}\eta =\partial _{\eta }f-e^i\partial _if$
from which we find $e^i\partial _if=\partial _{\eta }f-{\rm d}f/{\rm
d}\eta $. Then, straightforward calculations show that
\begin{eqnarray}
\label{Egi}
& & \frac{E_{_{\rm D}}}{E_{_{\rm E}}} =
\frac{a(\eta _{\rm lss})}{a(\eta _0)}
\biggl\{ 1
+\frac{1}{4}(\delta _{\gamma })_{\rm g}(\eta _{\rm lss},x^i_{_{\rm E}})
-\frac{1}{3}(\delta _{\rm b})_{\rm g}(\eta _0,x^i_{_{\rm D}})
+\Phi (\eta _{\rm lss},x^i_{_{\rm E}})-\Phi(\eta _0,x^i_{_{\rm D}})
\nonumber \\
& & +e^i\partial _iv^{\rm (gi)}(\eta _{0},x^i_{_{\rm D}})
-e^i\partial _iv^{\rm (gi)}(\eta _{\rm lss},x^i_{_{\rm E}})
+\int _{\eta _{\rm lss}}^{\eta _0}{\rm d}\tau 
\biggl[\Phi'+\Psi '-\frac{1}{2}
h_{ij}^{\rm (T)}{}'e^ie^j\biggr]\biggr\}\, .
\end{eqnarray}
We have thus proved the gauge invariance of the ratio $E_{_{\rm
D}}/E_{_{\rm E}}$~\cite{panek}.

\par

The Sachs-Wolfe effect is frequency independent. This means that the
shape of the black body is preserved at the perturbed level and this
is why a perturbed temperature is still a meaningful concept. If we
define $\delta T/T\equiv [\delta T_{_{\rm D}}-T_{_{\rm D}}]/T_{_{\rm
D}}$ with $T_{_{\rm D}}=T_{_{\rm E}}a(\eta _{\rm lss})/a(\eta _0)$, we
arrive at the final form of the Sachs-Wolfe effect, namely
\begin{equation}
\label{sw}
\frac{\delta T}{T}=
\biggl(\frac{\delta T}{T}\biggr)^{\rm (D)}
+\biggl(\frac{\delta T}{T}\biggr)^{\rm (S)}
+\biggl(\frac{\delta T}{T}\biggr)^{\rm (T)}\, ,
\end{equation}
with,
\begin{eqnarray}
\biggl(\frac{\delta T}{T}\biggr)^{\rm (D)} &=& 
e^i\partial _iv^{\rm (gi)}(\eta _{0},x^i_{_{\rm D}})
\nonumber \\
\biggl(\frac{\delta T}{T}\biggr)^{\rm (S)} &=&
\frac{1}{4}(\delta _{\gamma })_{\rm g}(\eta _{\rm lss},x^i_{_{\rm E}})
+\Phi (\eta _{\rm lss},x^i_{_{\rm E}})
-e^i\partial _iv^{\rm (gi)}(\eta _{\rm lss},x^i_{_{\rm E}})
\nonumber \\
& & +\int _{\eta _{\rm lss}}^{\eta _0}{\rm d}\tau
\left(\Phi'+\Psi '\right)\, ,
\nonumber \\
\biggl(\frac{\delta T}{T}\biggr)^{\rm (T)} &=&
-\frac{1}{2}\int _{\eta _{\rm lss}}^{\eta _0}{\rm d}\tau
\, \frac{\partial }{\partial \eta }h_{ij}^{\rm (T)}e^ie^j\, .
\end{eqnarray}
Several comments are in order here. Firstly, we have discarded the
terms $\Phi(\eta _{0},x^i_{_{\rm D}})$ and $\delta _{\rm b}^{\rm
(gi)}(\eta _{0},x^i_{_{\rm D}})/3$ since they do not depend on the
vector $e^i$. Secondly, the first term $[\delta T/T]^{\rm (D)}$ has
its $e^i$ dependence fixed. This is just the dipole term due to our
motion with respect to the frame of the CMBR. Thirdly, the other terms
are genuine fluctuations of primordial origin. As already mentioned,
they have been discovered in $1992$ by the COBE satellite.

\par

Finally, let us conclude this section by establishing the expression
of the Sachs-Wolfe effect due to density perturbations on large
scales. On these scales, the Doppler term is negligible. The
integrated Sachs-Wolfe effect is also negligible because, on
superhorizon scales, the Bardeen potential is approximatively
constant, hence its derivative vanishes (see before). Therefore, only
the first two terms remain. One can show that they combine such that
\begin{equation}
\label{swlarge}
\biggl(\frac{\delta T}{T}\biggr)^{\rm (S)}\simeq \frac{1}{3}
\Phi (\eta _{\rm lss},x^i_{_{\rm E}})\, .
\end{equation}
This equation permits to compute the angular power spectrum in the
COBE regime, i.e. for large angular scales.

\section{Quantization of Cosmological Perturbations}
\label{sec:4}

We start this section with a discussion of the quantization of a free
scalar field. This constitutes the prototype of methods used in the
sequel for the cosmological perturbations.

\subsection{Quantization of a Free Scalar Field}
\index{quantization!scalar field}

We consider the question of quantizing a (massless) scalar field in
curved space-time. The starting point is the following action
\begin{equation}
S=-\frac{1}{c}\int {\rm d}^4x \sqrt{-g}g^{\mu \nu}\frac12 \partial
_{\mu }\Phi \partial _{\nu }\Phi \, ,
\end{equation}
which, in a FLRW Universe, reads
\begin{equation}
S=\frac{1}{2c}\int {\rm d}^4x a^2(\eta )\left(\phi '^2 -\delta
^{ij}\partial _i\Phi \partial _j\Phi \right)\, .
\end{equation} 
It follows immediately that the conjugate momentum to the scalar 
field can be expressed as
\begin{equation}
\Pi (\eta ,\vec{x})=\frac{a^2}{c}\Phi '(\eta ,\vec{x})\, .
\end{equation}
It is convenient to Fourier expand the field $\Phi (\eta ,\vec{x})$
over the basis of plane waves (therefore, here, we use explicitly the
fact that the spacelike hypersurfaces are flat). This gives
\begin{equation}
\label{decompgw}
\Phi (\eta ,\vec{x})=\frac{1}{a(\eta )} \frac{1}{(2\pi )^{3/2}}\int
{\rm d}\vec{k} \mu _{\vec{k}}(\eta ){\rm e}^{i\vec{k}\cdot \vec{x}}\,
.
\end{equation} 
We have chosen to define the Fourier component with a factor $1/a(\eta
)$ for future convenience. Since the scalar field is real, this last
relation allows us to write $\mu _{\vec{k}}^*=\mu _{-\vec{k}}$. The
next step consists in inserting the expression of $\Phi (\eta
,\vec{x})$ into the action. This gives
\begin{eqnarray}
S &=& \frac{1}{2c} \int {\rm d}\eta \int _{R^{3+}}{\rm d}^3\vec{k}
\biggl[\mu _{\vec{k}}'{}^*\mu _{\vec{k}}'+ \mu _{\vec{k}}'\mu
_{\vec{k}}'{}^* -2\frac{a'}{a}\left(\mu _{\vec{k}}'\mu
_{\vec{k}}^*+\mu _{\vec{k}}'{}^*\mu _{\vec{k}}\right) \nonumber \\ & &
+\biggl(\frac{a'{}^2}{a^2}-k^2\biggr) \left(\mu _{\vec{k}}\mu
_{\vec{k}}^*+ \mu _{\vec{k}}^*\mu _{\vec{k}}\right)\biggr]\, .
\end{eqnarray}
Notice that the integral over the wavenumbers is calculated in half of
the space in order to sum over independent variables only. Equipped
with the Lagrangian in the momentum space (that, in the following, we
denote by $\bar{\cal L}$), we can now go to the Hamiltonian
formalism. The conjugate momentum to $\mu _{\vec{k}}$ is defined by
the formula
\begin{equation}
p_{\vec{k}}\equiv \frac{\delta \bar{{\cal L}}}{\delta \mu
_{\vec{k}}'{}^*} =\frac{1}{c}\left(\mu _{\vec{k}}'-\frac{a'}{a} \mu
_{\vec{k}}\right)\, .
\end{equation}
One can check that the definitions of the conjugate momenta in the
real and Fourier spaces are consistent in the sense that they are
linked by the (expected) expression
\begin{equation}
\Pi (\eta ,\vec{x})=\frac{a(\eta )}{(2\pi )^{3/2}}\int {\rm
d}{\vec{k}}p_{\vec{k}}{\rm e}^{i{\vec{k}}\cdot {\vec{x}}}\, .
\end{equation}
We see that the definition of the conjugate momentum $p_{\vec{k}}$ as
the derivative of the Lagrangian in the Fourier space with respect to
$\mu _{\vec{k}}'{}^*$ and not to $\mu _{\vec{k}}'$ is consistent with
the expression of the momentum in the real space. Otherwise the
momentum $\Pi (\eta ,\vec{x})$ in real space would have been expressed
in terms of $p_{\vec{k}}^*$ instead of $p_{\vec{k}}$.

\par

One can also check that the Lagrangian leads to the correct equation
of motion. Since we have $\delta \bar{\cal L}/\delta \mu
_{\vec{k}}^*=1/(2c)[-2{\cal H} \mu _{\vec{k}}'{}+2({\cal H}^2-k^2)\mu
_{\vec{k}}'{}]$, the Euler-Lagrange equation $ {\rm d}[\delta
\bar{\cal L}/\delta \mu _{\vec{k}}'{}^*]/{\rm d}\eta -\delta \bar{\cal
L}/\delta \mu _{\vec{k}}^*=0$ reproduces the correct equation of
motion for the variable $\mu _{\vec{k}}$, namely
\begin{equation}
\label{eq:motion}
\frac{{\rm d}^2\mu _{\vec{k}}}{{\rm d}\eta ^2}+
\left(k^2-\frac{a''}{a}\right)\mu _{\vec{k}}=0\, ,
\end{equation}
which is indeed the well-known result.

\par

We are now in a position where we can go to the Hamiltonian
formalism. The Hamiltonian density, $\bar{\cal H}$, is defined by
\begin{equation}
\bar{\cal H}\equiv p_{\vec{k}}\mu _{\vec{k}}'{}^*+p_{\vec{k}}^*\mu
_{\vec{k}}' -\bar{\cal L}\, ,
\end{equation}
and we obtain
\begin{equation}
\label{Hamil}
\bar{\cal H}=c\left(p_{\vec{k}}p_{\vec{k}}^*+\frac{k^2}{c^2}\mu
_{\vec{k}} \mu _{\vec{k}}^*\right)+\frac{a'}{a}\left(p_{\vec{k}}\mu
_{\vec{k}}^* +p_{\vec{k}}^*\mu _{\vec{k}}\right)\, .
\end{equation}
One can check that the Hamilton equations
\begin{equation}
\frac{{\rm d}\mu _{\vec{k}}^*}{{\rm d}\eta}
=\frac{\partial \bar{\cal H}}{\partial p_{\vec{k}}}
=cp_{\vec{k}}^*+\frac{a'}{a}\mu _{\vec{k}}^*\, ,
\quad
\frac{{\rm d}p_{\vec{k}}^*}{{\rm d}\eta}
=-\frac{\partial \bar{\cal H}}{\partial \mu _{\vec{k}}}
=-\frac{a'}{a}p_{\vec{k}}^*-\frac{k^2}{c}\mu _{\vec{k}}^*\, ,
\end{equation}
lead to the correct equation of motion given by Eq.~(\ref{eq:motion}).

\par

As a preparation to canonical quantization, we now introduce the
normal variable $\alpha _{\vec{k}}$~\cite{cohen} defined by
\begin{equation}
\alpha _{\vec{k}}(\eta )\equiv N(k)\mu _{\vec{k}}+
ic\frac{M(k)}{k}p_{\vec{k}}\, ,
\end{equation}
where, for the moment, the functions $N(k)$ and $M(k)$ are free but
will be specified later on. In terms of the normal variables, the
scalar field and its conjugate momentum can be expressed as
\begin{eqnarray}
\Phi (\eta ,\vec{k}) &=& \frac{1}{a(\eta )} \frac{1}{(2\pi )^{3/2}}
\int \frac{{\rm d}{\vec{k}}}{2N(k)}\, 
\left[\alpha _{\vec{k}}(\eta ){\rm e}^{i{\vec{k}}\cdot {\vec{x}}}
+\alpha _{\vec{k}}^*(\eta ){\rm e}^{-i{\vec{k}}\cdot {\vec{x}}}\right]
\, , \\
\Pi (\eta ,\vec{x}) &=& 
\frac{a(\eta )}{(2\pi )^{3/2}}\int {\rm d}{\vec{k}}\frac{k}{2icM(k)}
\left[\alpha _{\vec{k}}(\eta ){\rm e}^{i{\vec{k}}\cdot {\vec{x}}}
-\alpha _{\vec{k}}^*(\eta ){\rm e}^{-i{\vec{k}}\cdot {\vec{x}}}\right]
\, .
\end{eqnarray}
We are now ready to quantize the system. So far, we were dealing with
a relativistic field theory and only the constant $c$ appeared in the
equations. Now, the constant $\hbar $, which fixes the amplitude of
the fluctuations shows up. Concretely, the quantization is carried out
by requiring that $\Phi (\eta ,\vec{x})$ and $\Pi (\eta ,\vec{x})$
become quantum operators satisfying the usual commutation relation,
namely
\begin{equation}
\label{commut}
[\hat{\Phi }(\eta ,\vec{x}),\hat{\Pi }(\eta ,\vec{y})] =i\hbar \delta
^3(\vec{x}-\vec{y})\, .
\end{equation}
The normal variable $\alpha _{\vec{k}}(\eta )$ is promoted to an
operator $c_{\vec{k}}(\eta )$. We choose the commutation relation to
be $[c_{\vec{k}}(\eta ), c_{\vec{p}}^{\dagger}(\eta )]=C\delta
({\vec{k}}- {\vec{p}})$. In the last expression, $C$ is a free
dimensionless constant. Notice that the commutation relation is
time-independent. Then, the expressions of $N(k)$ and $M(k)$ are fully
determined. Let us see in more details how the calculation
proceeds. The commutator is given by
\begin{eqnarray}
[\hat{\Phi }(\eta ,\vec{x}),\hat{\Pi }(\eta ,\vec{y})] &=&
\frac{iC}{4c(2\pi )^3} \int {\rm d}^3{\vec{k}}
\frac{k}{N(k)M(k)}\left[ {\rm e}^{i\vec{k}(\vec{x}-\vec{y})} +{\rm
e}^{-i\vec{k}(\vec{x}-\vec{y})} \right]\, .
\end{eqnarray}
We see that, in order to produce a Dirac function $\delta
^3(\vec{x}-\vec{y})$ which is necessary in order to reproduce the
relation given by Eq.~(\ref{commut}) by integration of the
exponentials, the term $k/(NM)$ must be $k$-independent. The link
between the functions $N(k)$ and $M(k)$ is therefore determined. Let
us call $D$ the term $k/(NM)$. Then the result reads
\begin{equation}
[\hat{\Phi } (\eta ,\vec{x}),\hat{\Pi }(\eta ,\vec{y})] =
\frac{iC}{4c}\times 2D\delta ^3(\vec{x}-\vec{y})\, .
\end{equation}
As a consequence we have $CD=2\hbar c$. As expected, the normalization
is given by a combination of $\hbar $ and $c$. In the following, we
will adopt the convenient choice $C=1$.

\par

Everything has been fixed but the function $N(k)$. This function is
chosen by means of the following considerations. The energy of a
scalar field is given by the formula
\begin{eqnarray}
\hat{E} &=& \int {\rm d}^3{\vec{x}}\sqrt{-{}^{(3)}g}\hat{\rho } =\int
{\rm d}^3{\vec{x}}\sqrt{-{}^{(3)}g}
\frac{1}{2a^2}\left(\hat{\Phi}'{}^2+\delta ^{ij}\partial _i \hat{\Phi}
\partial _j \hat{\Phi }\right) \\ &=& \int {\rm
d}^3{\vec{x}}\sqrt{-{}^{(3)}g} \frac{1}{2a^2}\left[\left(\frac{c}{a^2}
\hat{\Pi }\right)^2+\delta ^{ij}\partial _i \hat{\Phi }\partial _j
\hat{\Phi }\right]\, ,
\end{eqnarray}
where we have used the expression of the conjugate momentum. In this
expression, the determinant of the metric is the determinant of the
spatial part of the metric (including the factor $a$). We can now
insert the expression of the operators $\hat{\Phi }$ and $\hat{\Pi }$
in the above equation giving $\hat{E }$. One finds
\begin{eqnarray}
\label{eq:energy}
\hat{E }&=& \frac{1}{2a}
\int {\rm d}^3{\vec{k}}\frac{1}{4}
\biggl[\frac{k^2}{M^2(k)}\left(-c_{\vec{k}}c_{-{\vec{k}}}
+c_{\vec{k}}c_{{\vec{k}}}^{\dagger }+c_{\vec{k}}^{\dagger }
c_{{\vec{k}}}-c_{\vec{k}}^{\dagger }c_{-{\vec{k}}}^{\dagger }\right)
\nonumber \\
& & +\frac{k^2}{N^2(k)}\left(c_{\vec{k}}c_{-{\vec{k}}}
+c_{\vec{k}}c_{{\vec{k}}}^{\dagger }+c_{\vec{k}}^{\dagger }
c_{{\vec{k}}}
+c_{\vec{k}}^{\dagger }c_{-{\vec{k}}}^{\dagger} \right)\biggr]\, .
\end{eqnarray}
Our criterion is to put ``half of a quanta in each
mode''. Technically, this means that we would like the energy to take
the following suggestive form
\begin{equation}
\label{energy}
\hat{E} =\int {\rm d}^3{\vec{k}}\frac{\hbar \omega (\eta )}{2}
\left(c_{\vec{k}}c_{{\vec{k}}}^{\dagger }
+c_{{\vec{k}}}^{\dagger }c_{\vec{k}}\right)\, ,
\end{equation}
where $\omega (\eta )=kc/a(\eta )$ is the physical frequency. We see
that the only way to cancel the unnecessary terms in
Eq.~(\ref{eq:energy}) is to have $N(k)=M(k)$. Together with the
relation established previously, $D=2\hbar c$, this gives
$N^2(k)=k/(2\hbar c)$. As a consequence, The scalar field operator now
reads
\begin{equation}
\hat{\Phi }(\eta ,\vec{x})=\frac{\sqrt{\hbar c}}{a(\eta )}
\frac{1}{(2\pi )^{3/2}}\int \frac{{\rm d}{\vec{k}}}{\sqrt{2k}}
\left[c_{\vec{k}}(\eta ){\rm
e}^{i{\vec{k}}\cdot {\vec{x}}} +c_{\vec{k}}^{\dagger }(\eta ){\rm
e}^{-i{\vec{k}}\cdot {\vec{x}}}\right] \, .
\end{equation} 
Everything is now fixed. The expression of the scalar field operator
contains no unspecified factor. Even the amplitude is fixed and is
given by the factor $\sqrt{\hbar c}$.

\par

We can now calculate the Hamiltonian operator. Using Eq.~(\ref{Hamil})
one obtains
\begin{equation}
\hat{{\bf H}}=\frac{1}{2}\int _{R^3}{\rm d}^3\vec{k} \left[\hbar
k\left(c_{\vec{k}}c_{\vec{k}}{}^{\dagger } +c_{-{\vec{k}}}{}^{\dagger
}c_{-{\vec{k}}}\right ) -i\hbar
\frac{a'}{a}\left(c_{\vec{k}}c_{-{\vec{k}}} +c_{-{\vec{k}}}{}^{\dagger
}c_{{\vec{k}}}{}^{\dagger }\right )\right]\, ,
\end{equation}
where it is important to notice that the integral is calculated in
$R^3$ and not in $R^{3+}$. Let us analyze this Hamiltonian. The first
term is the standard one and represents a collection of harmonic
oscillators. The most interesting part is the second term. This term
is responsible for the quantum creation of particles in curved
spacetime. It can be viewed as an interacting term between the scalar
field and the classical background. The coupling function $ia'/a$ is
proportional to the derivative of the scale factor and therefore
vanishes in flat spacetime. From the structure of the interacting
term, i.e. in particular the product of two creation operators for the
mode ${\vec k}$ and $-{\vec{k}}$, we can also see that we have
creation of pairs of quanta with opposite momenta during the
cosmological expansion.

\par

We can now calculate the time evolution of the quantum operators (we
are here in the Heisenberg picture). Everything is known if we can
determine what the temporal behavior of the creation and annihilation
behavior is. The temporal behavior is given by the Heisenberg
equations which read
\begin{equation}
\frac{{\rm d}c_{{\vec{k}}}}{{\rm d}\eta
}=-\frac{i}{\hbar}[c_{{\vec{k}}},\hat{{\bf H}}]\, , \quad \frac{{\rm
d}c_{{\vec{k}}}{}^{\dagger }}{{\rm d}\eta }
=-\frac{i}{\hbar}[c_{{\vec{k}}}{}^{\dagger },\hat{{\bf H}}]\, .
\end{equation}
Inserting the expression of the Hamiltonian derived above, we 
arrive at the equations
\begin{equation}
\frac{{\rm d}c_{{\vec{k}}}}{{\rm d}\eta }=kc^s_{{\vec{k}}}
+i\frac{a'}{a}c_{-{\vec{k}}}{}^{\dagger }\, ,\quad 
\frac{{\rm d}c_{{\vec{k}}}{}^{\dagger }}{{\rm d}\eta }
=-kc^s_{{\vec{k}}}{}^{\dagger }
-i\frac{a'}{a}c^s_{-{\vec{k}}}\, .
\end{equation}
This system of equations can be solved by means of a Bogoliubov
transformation and the solution can be written as
\begin{eqnarray}
c_{\vec{k}}(\eta ) &=& u_k(\eta )c_{\vec{k}}(\eta _{\rm ini})
+v_k(\eta )c_{-{\vec{k}}}{}^{\dagger }(\eta _{\rm ini})\, ,
\\
c_{\vec{k}}{}^{\dagger }(\eta ) &=& u_k^*(\eta )
c_{\vec{k}}{}^{\dagger }(\eta _{\rm ini})
+v_k^*(\eta )c_{-{\vec{k}}}{}(\eta _{\rm ini})\, ,
\end{eqnarray}
where $\eta _{\rm ini}$ is a given initial time and where the
functions $u_k(\eta )$ and $v_k(\eta )$ satisfy the equations
\begin{equation}
\label{eqsuv}
i\frac{{\rm d}u_k(\eta )}{{\rm d}\eta }=ku_k(\eta )
+i\frac{a'}{a}v_k^*(\eta )\, ,\quad 
i\frac{{\rm d}v_k(\eta )}{{\rm d}\eta }=kv_k(\eta )
+i\frac{a'}{a}u_k(\eta )\, .
\end{equation}
In addition, these two functions must satisfy $\vert u_k\vert ^2
-\vert v_k\vert ^2=1$ such that the commutation relation between the
creation and annihilation operators is preserved in time. A very
important property is the initial values of the two functions are
fixed and, from the Bogoliubov transformation, read
\begin{equation}
\label{iniquant}
u_k(\eta _{\rm ini})=1\, , \quad v_k(\eta _{\rm ini})=0\, .
\end{equation}

\par

At this point, the next move is to establish the link between the
formalism exposed above and the classical picture. For this purpose,
it is interesting to establish the equation of motion obeyed by the
function $u_k+v_k^*$. Straightforward manipulations from
Eqs.~(\ref{eqsuv}) lead to
\begin{equation}
\label{eomquant}
\left(u_k+v_k^*\right)''
+\biggl(k^2-\frac{a''}{a}\biggr)\left(u_k+v_k^*\right)=0\, .
\end{equation}
Therefore, the function $u_k+v_k^*$ obeys the same equation as the
variable $\mu _{\vec{k}}$. This is to be expected since, using the
Bogoliubov transformation, the scalar field operator can be re-written
as
\begin{eqnarray}
\hat{\Phi }(\eta ,{\bf x}) &=& \frac{\sqrt{\hbar c}}{a(\eta )}
\frac{1}{(2\pi )^{3/2}} \int \frac{{\rm d}{\vec{k}}}{\sqrt{2k}}
\biggl[\left(u_k+v_k^*\right)(\eta )c_{\vec{k}}(\eta _{\rm ini}) {\rm
e}^{i{\vec{k}}\cdot {\vec{x}}} \nonumber \\ & &
+\left(u_k^{*}+v_k\right)(\eta )c_{\vec{k}}^{\dag}(\eta _{\rm ini})
{\rm e}^{-i{\vec{k}}\cdot{\vec{x}}}\biggr]\, .
\end{eqnarray}
If we are given a scale factor, we can now calculate completely the
time evolution of the perturbations by means of the formalism
presented above. Let us stress again that the quantization procedure
has completely fixed the overall amplitude of the field. Indeed, the
field is normalized to $\sqrt{\hbar c}$ while the ``mode function''
$u_k+v_k^*$ has initially an amplitude of one.

\par

Let us now calculate the two-point correlation function in the 
vacuum state. One gets
\begin{equation}
\left\langle 0\left \vert \hat{\Phi} (\eta ,\vec{x})\hat{\Phi }(\eta
,\vec{x}+\vec{r})\right \vert 0\right \rangle =\frac{\hbar c}{4\pi
^2}\int _0^{+\infty }\frac{{\rm d}k}{k}\frac{\sin kr}{kr} k^2\left
\vert \frac{u_k+v_k^*}{a(\eta )}\right \vert ^2\, .
\end{equation}
If we assume that the scale factor is given by a power-law of the
conformal time, $a(\eta )=\ell _0(-\eta )^{1+\beta }$, where $\beta\le
-2$ is a free a parameter and $\ell _0$ a constant with the dimension
of a length, then the solution of Eq.~(\ref{eomquant}) with the initial 
conditions given by Eqs.~(\ref{iniquant}) reads
\begin{equation}
\label{solhankel}
\left(u_k+v_k^*\right)(\eta )=\sqrt{\frac{\pi }{2}}{\rm e}^{i(k\eta
_{\rm ini} -\pi \beta /2)}\sqrt{-k\eta }H^{(1)}_{-\beta
-1/2}\left(-k\eta \right)\, ,
\end{equation}
where $H^{(1)}$ is a Hankel function of first kind. From, this
solution, it is easy to calculate the spectrum on large angular scales
($k\eta \to 0$)
\begin{equation}
\frac{\hbar c}{4\pi ^2} k^2\left \vert \frac{u_k+v_k^*}{a(\eta
)}\right \vert ^2=\frac{\hbar c}{4\pi ^2}\frac{f(\beta )}{\ell _0^2}
k^{4+2\beta }\, ,
\end{equation}
where $f(\beta )\equiv \pi ^{-1}\left[2^{-1-\beta }\Gamma \left(-\beta
-1/2 \right)\right]^2$. In particular, if $\beta =-2$, this case
corresponding to de Sitter spacetime for which the Hubble constant is
strictly constant, one has $\ell _0=c/H_{_{\rm inf}}$ and the spectrum
reads
\begin{equation}
\frac{\hbar}{c}\left(\frac{H_{_{\rm inf}}}{2\pi }\right)^2\, 
\end{equation}
i.e. is scale-invariant (which means that it does not depend on the
wavenumber). Of course, if $\beta \neq -2$ then the spectrum is scale
dependent.

\par

The above result leads us to a first attempt to quantize cosmological
perturbations~\cite{KT}. Using Eq.~(\ref{spectrumzeta}) and taking
into account the fact that $\omega _{_{\rm inf}}\simeq -1$ and ${\cal
H}/\varphi '\simeq \kappa /(2\epsilon )$ (recall that $\epsilon $ is
the first slow-roll parameter) , one obtains for the spectrum of the
``tracer'' $\zeta _k$
\begin{equation}
P_{\zeta}\equiv k^3\zeta _k^2\simeq \frac{\kappa }{2\epsilon
}k^3\left[\delta \varphi _k^{\rm (gi)} \right]^2\, .
\end{equation}
The question is now how should we calculate $\left[\delta \varphi
_k^{\rm (gi)} \right]^2$? Historically, the idea was to consider that
the matter fluctuations (i.e. fluctuations in the scalar field) are
quantized while the fluctuations in the gravitational field remain
classical. Based on this guess, one can used the trick which consists
in replacing
\begin{equation} 
\label{trick}
\left[\delta \varphi ^{\rm (gi)} \right]^2 \rightarrow \left\langle
0\left\vert \left[\delta \hat{\varphi }^{\rm (gi)} \right]^2 \right
\vert 0\right \rangle \,
\end{equation}
or, in the Fourier space, $\left[\delta \varphi _k^{\rm (gi)}
\right]^2 \rightarrow \hbar H_{_{\rm inf}}^2/(4\pi ^2 c)$. This gives for 
the spectrum of density perturbations
\begin{equation}
\label{spectracer}
P_{\zeta }\equiv k^3\zeta _k^2 \simeq \frac{\hbar
G}{c^5}\frac{H_{_{\rm inf}}^2}{\pi \epsilon }\, .
\end{equation}
As expected the three fundamental constants, $G$, $c$ and $\hbar $
participate to the final expression. We have kept them in order to be
able to trace back their origin. The combination which appears here is
the Planck time squared as it has to be since $\zeta_k$ is a
dimensionless quantity. In natural units, the above spectrum is just
$H_{_{\rm inf}}^2/(\pi \epsilon m_{_{\rm Pl}}^2)$. Several remarks are
in order at this point. Firstly, as we will see, this trick provides
us with the exact result. Secondly, it seems is that there is no way
to rigorously justify the replacement~(\ref{trick}). The reason is
that matter, i.e. $\delta T_{\mu \nu}$, is treated
quantum-mechanically, while geometry, i.e. $\delta G_{\mu \nu}$, is
still considered to be classical, despite the fact that both are
linked by the perturbed Einstein equations, $\delta G_{\mu \nu}=\kappa
\delta T_{\mu \nu}$. One could think that a semi-classical equation
\begin{equation}
\label{semi}
\delta G_{\mu \nu }=\kappa \left \langle 0\left \vert \delta
\hat{T}_{\mu \nu }\right \vert 0\right \rangle \, ,
\end{equation}
could do the job but in fact one easily realizes that this cannot be
the case because $\delta T_{\mu \nu}$ being linear in $\delta \varphi
$ (it is of course quadratic in the scalar fields, but at linear order
we have terms like $\varphi ' \delta \varphi '$), we have in fact
$\left \langle 0\left \vert \delta \hat{T}_{\mu \nu }\right \vert
0\right \rangle =0$ due to $\left \langle 0\left \vert \delta
\hat{\varphi }\right \vert 0\right \rangle =0$.  Therefore,
Eq.~(\ref{semi}) is in fact inconsistent in the present
context. Thirdly, it would be dangerous to base the physical
interpretation of Eq.~(\ref{spectracer}) on the above method arguing
that it gives the correct result. Here, we emphasize that a convincing
physical interpretation should be based on a consistent framework. We
can try the following analogy. The correct equation for the energy
levels of an Hydrogen atom, $E_n\propto 1/n^2$, has been obtained for
the first time by means of the so-called Bohr's model. This model was
developed before a consistent framework for Quantum Mechanics become
available. But, it is clear that, today, nobody would try to use
Bohr's framework to interpret the formula for $E_n$. We are of the
opinion that the situation for the cosmological perturbations is
similar. Fourthly, the correct way to proceed is to treat the
fluctuations in the geometry and in the scalar field on an equal
footing. This amounts to ``quantize'' both sides of the Einstein
equations and to write~\cite{MFB}
\begin{equation}
\delta \hat{G}_{\mu \nu}=\kappa \delta \hat{T}_{\mu \nu}\, .
\end{equation}
The consequence is of course of utmost importance: the metric operator
$h_{\mu \nu}$ should now be considered as a quantum operator, $h_{\mu
\nu}\rightarrow \hat{h}_{\mu \nu}$. In other words, we have now to
deal with the quantum-mechanical nature of the gravitational field,
i.e. with quantum gravity (at the linearized level). We now turn to
this question.

\subsection{Quantization of Density Perturbations}
\index{quantization!density perturbations}

The total action of the system is given by

\begin{equation}
S=-\frac{c^3}{16\pi G}\int {\rm d}^4x\sqrt{-g}R -\frac{1}{c}\int {\rm
d}^4x\sqrt{-g} \biggl[\frac{1}{2}g^{\mu \nu}\partial _{\mu} \varphi
\partial _{\nu}\varphi +V(\varphi )\biggr]\, .
\end{equation}
If we perturb this action up to second order in the metric
perturbations and in the scalar field fluctuations (this is necessary
if we want the variation of this action to reproduce the first order
equations of motion) one finds, despite a very long and tedious
calculation, that the result is delightfully simple, namely~\cite{MFB}
\begin{equation}
 \label{scaleac} {}^{(2)}\delta S={1 \over 2c} \int {\rm d}^4x
\biggl[(v')^2- \delta^{i j} \partial _iv\partial _jv+{{z_{_{\rm S}}''}
\over z_{_{\rm S}}} v^2 \biggr] \, ,
\end{equation}
with
\begin{equation} 
v(\eta ,\vec{x}) \equiv a \left[ \delta \varphi ^{\rm (gi)} +
{{\varphi '} \over {\cal H}} \Phi \right]\, .
\end{equation}
This is nothing but the action for a scalar field with a
time-dependent mass. The constant $G$ does not appear explicitly in
the above action because it has been absorbed via the background
Einstein equations. It is interesting to notice that the natural
variable is not $\Phi $ neither $\zeta $ but $v$. It is not a surprise
that the system is characterized by a single quantity since
gravitational fluctuations are described by $\Phi $ and matter
fluctuations by $\delta \varphi ^{\rm (gi)}$ but are linked by the
perturbed Einstein equations. Therefore, only one degree of freedom is
left. The link between $v$ and the ``tracer'' $\zeta $ is given by
\begin{equation}
\zeta =\sqrt{\frac{\kappa }{2}} \frac{v}{a\sqrt{\epsilon }}\, .
\end{equation}
Finally, the quantity $z_{_{\rm S}}$ is given by $z_{_{\rm S}}
=\sqrt{\kappa /2}a\varphi '/{\cal H}=a\sqrt{\epsilon }$ because, from
the background Einstein equations, one has $\kappa (\varphi ')^2
=2{\cal H}^2\epsilon $.

\par

At this point, the procedure of quantization follows exactly the one
presented in the last subsection. The quantity $v(\eta ,\vec{x})$
becomes a quantum operator $\hat{v}(\eta ,\vec{x})=a\left[\hat{\delta
\varphi ^{\rm (gi)}}+\left(\varphi '/{\cal H}\right)\hat{\Phi }\right]
$, the expression of which can be written as
\begin{eqnarray}
\label{candecomp}
\hat{v}(\eta ,\vec{x}) &=& \frac{\sqrt{\hbar c}}{(2\pi )^{3/2}}\int
\frac{{\rm d}^3{\vec{k}}}{\sqrt{2k}} \biggl[(u_k+v_k^*)(\eta )
c_{\vec{k}}(\eta _{\rm ini}){\rm e}^{i {\vec{k}} \cdot {\vec{x}}}
\nonumber \\
& & +(u_k^*+v_k)(\eta ) c_{\vec{k}}^{\dagger }(\eta _{\rm ini})
{\rm e}^{-i{\vec{k}} \cdot {\vec{x}}}\biggr] \, .
\end{eqnarray} 
As announced, the fluctuations of the metric tensor are now quantized:
technically, the Bardeen potential $\Phi (\eta ,\vec{k})$ is now a
quantum operator $\hat{\Phi }(\eta ,\vec{x})$ which explicitly appears
into the expression of $\hat{v}(\eta ,\vec{x})$. Notice also that the
uncertainty principle has completely fixed the overall amplitude of
the quantum perturbations since the initial conditions are fixed by
$u_k(\eta _{\rm ini})=1$ and $v_k(\eta _{\rm ini})=0$. The equation of
motion reads
\begin{equation} 
\label{scaleeq} 
\left(u_k+v_k^*\right)'' + \left( k^2 - \frac{z_{_{\rm S}}''}{z_{_{\rm
S}}} \right) \left(u_k+v_k^*\right)=0 \, .
\end{equation}
The physical meaning of the initial conditions are as follows:
initially, we choose the state which is empty of ``particles'' from
the point of view of a local comoving observer at the initial time
$\eta _{\rm ini}$. This state $|0\rangle$ is defined by $
c_{\vec{k}}\vert 0\rangle =0 $. Since, due to the time dependence of
the background, there is a nontrivial mixing between positive and
negative frequencies, this state is in general not the vacuum at later
times

\par

We are now in a position to calculate the power spectrum of the quantum 
operator $\hat{\zeta }$. One gets
\begin{eqnarray}
\left \langle 0\left \vert \hat{\zeta }(\eta ,{\vec{x}}) \hat{\zeta
}(\eta ,{\vec{x}}+{\vec{r}}) \right \vert 0\right \rangle &=&
\frac{\hbar c\kappa }{8\pi ^2z_{_{\rm S}}^2}\int _0^{+\infty }
\frac{{\rm d}k}{k}\frac{\sin kr}{kr}k^2 \vert u_k+v_k^*\vert ^2 \, ,
\end{eqnarray} 
from which we easily deduce the expression of the power spectrum
\begin{equation}
k^3P_{\zeta }=\frac{\hbar c\kappa }{8\pi ^2}k^2\biggl \vert 
\frac{u_k+v_k^*}{z_{_{\rm S}}(\eta )}\biggr\vert ^2\, .
\end{equation}
In order to compare this result with Eq.~(\ref{spectracer}), we can
evaluate the spectrum for power-law inflation where an exact solution
of the equation of motion is available. Indeed, for $a(\eta )=\ell
_0(-\eta )^{1+\beta }$, the function $\epsilon $ is a constant, hence
one has $z_{_{\rm S}}''/z_{_{\rm S}}=a''/a$. This means that the
Hankel function of Eq.~(\ref{solhankel}) is also solution of
Eq.~(\ref{scaleeq}). Then, straightforward calculations show that

\begin{equation}
k^3P_{\zeta }=\frac{1}{\pi \epsilon }\frac{\hbar G}{c^3\ell _0^2}
f\left(\beta \right)k^{2\beta +4}\, .
\end{equation}
If $\beta $ is close to $-2$ then $\ell _0\simeq c/H_{_{\rm inf}}$ and
one recovers exactly the result of Eq.~(\ref{spectracer}). As
expected, the Planck length ``naturally'' appears in the above result.

\subsection{Quantization of Gravitational Waves}
\index{quantization!gravitational waves}

The quantization of gravitational waves proceeds exactly along the
same lines as before. Therefore, in this subsection, we only review
briefly the main results. The starting point is the Einstein-Hilbert
action that we expand to the second order. One gets (in natural
units)~\cite{MFB}
\begin{equation}
{}^{(2)}\delta S=\frac{m_{_{\rm Pl}}^2}{64 \pi}\int
\left[(h^i{}_j)'(h^j{}_i)' -\partial _k(h^i{}_j) \partial
^k(h^j{}_i)\right]a^2(\eta ){\rm d}^4x \, .
\end{equation}
In fact, the action can be re-written as 
\begin{equation}
{}^{(2)}\delta S_2=-\frac{m_{_{\rm Pl}}^2}{16\pi}\sum _{s=+,\times }
\int {\rm d}^4x\frac{1}{2} g^{\mu \nu}\partial _{\mu }h^s\partial
_{\nu }h^s\, ,
\end{equation}
where the quantity $h^s(\eta ,\vec{x})$ is defined by
\begin{equation}
h^s(\eta ,\vec{x})\equiv \frac{1}{a(\eta )} \frac{1}{(2\pi
)^{3/2}}\sum _{s=+,\times }^2\int {\rm d}{\vec{k}}\mu _{_{\rm T}}
^s(\eta ,\vec{k}){\rm e}^{i{\vec{k}}\cdot {\vec{x}}}\, .
\end{equation} 
Therefore, the action of gravitational waves is equivalent to the
action of two decoupled scalar fields (corresponding to the two states
of polarization). One can then follow the method presented before. The
quantum perturbed metric operator can be written as
\begin{eqnarray}
\hat{h}_{ij}(\eta ,{\vec{x}}) &=& \frac{4\sqrt{\pi }}{m_{_{\rm Pl}}a(\eta
)} \frac{1}{(2\pi )^{3/2}} \sum _{s=+,\times }\int \frac{{\rm
d}{\vec{k}}}{\sqrt{2k}}p_{ij}^s({\vec{k}})
\biggl[(u_k^s+v_k^{s*})(\eta )c_{\vec{k}}^s(\eta _{\rm ini}){\rm
e}^{i{\vec{k}} \cdot {\vec{x}}}
\nonumber \\
& & +(u_k^{s*}+v_k^s)(\eta
)c_{\vec{k}}^{s\dag}(\eta _{\rm ini}) {\rm
e}^{-i{\vec{k}}\cdot{\vec{x}}}\biggr]\, ,
\end{eqnarray}
where the function $(u_n^s+v_n^s{}^*)(\eta )$ obeys
Eq.~(\ref{eomgw}). Finally, the two-point correlation function of the
perturbed metric operator can be expressed as
\begin{eqnarray}
& &\left\langle 0\left\vert \hat{h}_{ij}(\eta ,{\vec{x}})
\hat{h}^{ij}(\eta ,{\vec{x}}+{\vec{r}}) \right \vert 0 \right \rangle
\nonumber \\
& & =\frac{16}{\pi m_{_{\rm Pl}}^2a^2(\eta )} \int _0^{+\infty }\frac{{\rm
d}k}{k}\frac{\sin kr}{kr} k^2\vert u_k^s+v_k^s{}^*\vert ^2\, ,
\end{eqnarray}
from which we deduce that the power spectrum of the gravitational
waves is given by
\begin{equation}
k^3P_h(k,\eta )=\frac{16}{\pi m_{_{\rm Pl}}^2}
k^2\biggl\vert \frac{u_k^s+v_k^s{}^*}{a(\eta )}\biggr\vert ^2\, .
\end{equation}
If we had decided to keep the standard units, the factor $1/m_{_{\rm
Pl}}^2$ in the above result would have obviously read $\hbar G/c^3$,
i.e. the Planck length squared.

\subsection{The Power Spectra in the Slow-roll Approximation}

We have established the expression of the scalar and tensor power
spectra and calculated these quantities for power-law
inflation. However, as discussed at the beginning of this review
article, the most interesting physical situation occurs when the
slow-roll approximation is valid. As discussed previously, the only
thing we need to do in order to compute the spectrum is to solve the
equation of a parametric oscillator,
\begin{equation}
\label{parameq}
\mu ''+\left[k^2-U\left(\eta \right)\right]\mu = \mu
''+\left[k^2-\frac{z''}{z}\left(\eta \right)\right]\mu =0\, ,
\end{equation}
where $\mu $ is $u_k+v_k^*$ either for scalar or tensor perturbations
and the effective potential $z_{_{\rm S}}''/z_{_{\rm S}}$ or $a''/a$,
i.e. $z=z_{_{\rm S}}$ or $z=a(\eta )$. As already mentioned, on
subhorizon or superhorizon scales, this equation can be solved
regardless of the detailed form of the scale factor. The solutions
are $\exp\left(-ik\eta \right)$ and $z(\eta )+z(\eta )\int ^{\eta
}{\rm d}\tau z^{-2}(\tau )$ respectively. However, in order to obtain
a reliable solution, one also needs to know the form of the solution
in the regime $k^2\simeq U(\eta )$, that is to say when the
corresponding scales crossed out the horizon during inflation. 

\par

Let $N_*\left (\lambda \right)$ be the number of e-folds before the
end of inflation at which the scale $\lambda $ exits the horizon. We
have
\begin{equation}
\label{efoldexit}
N_*(\lambda )\simeq \ln \left(\frac{\lambda }{\ell _{_{\rm H}}}\right)
+\left[\log _{10}\left(\frac{H_{_{\rm inf}}}{m_{_{\rm Pl}}}\right)
-\log _{10}\left(\frac{T_{_{\rm RH}}}{m_{_{\rm Pl}}}\right)
+29\right]\times \ln 10\, .
\end{equation}
If we take the fiducial values $H_{_{\rm inf}}\simeq
10^{14}\mbox{GeV}$ and $T_{_{\rm RH}}\simeq M_{_{\rm inf}}\simeq
10^{16.5}\mbox{GeV}$ then $N_*\simeq 60$ for the Hubble scale today,
i.e. $\lambda \simeq \ell _{_{\rm H}}$, see also
Fig.~\ref{fig:scales}. A scale characterized by its wave-number $k$
corresponds to an angle $\theta $ on the celestial sphere of about
$k\simeq 1/(2\ell _{_{\rm H}}\theta )$. Given the present CMBR
experiments, this means that we probe in fact the scales $\ell _{_{\rm
H}}<\lambda <10^{-3}\ell _{_{\rm H}}$. The smallest scale in this
interval crossed out the horizon $\simeq 46$ e-folds before the end of
inflation. This means that the time taken by the the scales of
astrophysical interest today to cross the horizon during inflation
corresponds to $\Delta N\simeq 7$. Therefore, we need an accurate
description of the effective potential $U(\eta )$ only during $7$
e-folds, see Fig.~\ref{fig:scales}.

\par

In the slow-roll approximation, the effective potentials for scalar
and tensor read at linear order
\begin{equation}
U_{_{\rm S}}(\eta )=\frac{2+6\epsilon -3\delta }{\eta ^2}\, , \quad
U_{_{\rm T}}(\eta )=\frac{2+3\epsilon}{\eta ^2}\, .
\end{equation}
Moreover, the equations of motion for $\epsilon$ and $\delta$ can be
written as:
\begin{equation}
\label{eqmotionsrpara}
\frac{{\rm d}{\epsilon}}{H{\rm d}t}=\frac{{\rm d}{\epsilon}}{{\rm d}N}
=2\epsilon (\epsilon -\delta)\ , \quad \frac{{\rm d}{\delta }}{H{\rm
d}t}=\frac{{\rm d}{\delta }}{{\rm d}N} =2\epsilon (\epsilon -\delta
)-\xi \, .
\end{equation}
From these equations, one sees that typically ${\cal O}\left(\epsilon
^2\right) \Delta N\ll 1$ for $\Delta N\simeq 7$ and, therefore, the
slow-roll parameters can be considered as constant during the exit of
the physical modes. This simplifies the problem drastically since then
the equations of motion in the regime $k^2\simeq U(\eta )$ can be
solved in terms of Bessel functions whose orders depend on the
slow-roll parameters. A detailed calculation can be found in
Refs.~\cite{MS2} and, here, we just give the result
\begin{eqnarray}
\label{pssr}
k^3P_{\zeta } &=& {H^2\over \pi \epsilon m_{_{\rm Pl}}^2}
\left[1-2\left(C+1\right)\epsilon -2C\left(\epsilon -\delta \right)
-2\left(2\epsilon -\delta \right)\ln {k\over
k_*}\right]\, , 
\\ 
\label{pssrgw}
k^3P_{h} &=& {16 H^2\over \pi m_{_{\rm Pl}}^2}
\left[1-2\left(C+1\right)\epsilon -2\epsilon \ln {k\over k_*}\right]\,
,
\end{eqnarray}
where $C$ is a numerical constant, $C\simeq -0.73$ and $k_*$ a scale
called the ``pivot scale''. We see that the amplitude of the scalar
power spectrum is given by a scale-invariant piece, $H^2/(\pi \epsilon
m_{_{\rm Pl}}^2)$ that we had already guessed before, plus logarithmic
corrections the amplitude of which is controlled by the slow-roll
parameters, i.e. by the microphysics of inflation. It is important to
notice that $H$ is the value of the Hubble parameter during the $7$
e-folds where the scales of astrophysical interest crossed out the
horizon, see Fig.~\ref{fig:scales}. As already mentioned at the end of
Sec.~(2.4) this can be different from the value of the Hubble
parameter at the beginning of inflation. The above remarks are also
valid for tensor perturbations. The ratio of tensor over scalar is
just given by
\begin{equation}
\frac{k^3P_{h}}{k^3P_{\zeta }}=16\epsilon \, .
\end{equation}
This means that the gravitational are always sub-dominant and that,
when we measure the CMBR anisotropies, we essentially see the scalar
modes. This is rather unfortunate because this implies that one cannot
measure the energy scale of inflation since the amplitude of the
scalar power spectrum also depends on the slow-roll parameter
$\epsilon $. Only an independent measure of the gravitational waves
contribution could allow us to break this degeneracy. On the other
hand, the spectral indexes are given by
\begin{equation}
n_{_{\rm S}}=\frac{\ln k^3P_{\zeta }}{{\rm d}\ln k}\biggl\vert
_{k=k_*}=1 - 4\epsilon + 2\delta\, , \quad n_{_{\rm T}}=\frac{\ln
k^3P_h}{{\rm d}\ln k}\biggl\vert _{k=k_*}=-2\epsilon \, .
\end{equation}
As expected, the power spectra are always close to scale invariance
and the deviation from it is controlled by the magnitude of the two
slow-roll parameters. Finally, at the next-to-leading order there is
no running of the spectral indexes since they are in fact second order
in the slow-roll parameters.

\section{Comparison with Observations}
\label{sec:5}
\index{CMB anisotropy}

In this section, we briefly discuss the impact of the recent Wilkinson
Microwave Anisotropy Probe (WMAP) measurements on
inflation~\cite{wmap}. We have seen previously that the presence of
cosmological perturbations causes anisotropies in the CMBR (the
Sachs-Wolfe effect) and we have established the link between $\delta
T/T$ and the metric fluctuations, see Eqs.~(\ref{sw}) and
(\ref{swlarge}). The fact that the metric fluctuations are described
by a quantum operator has an immediate consequence: $\delta T/T $
should be considered as a quantum operator as well. It is convenient
to expand this operator on the celestial sphere, i.e. on the basis of
spherical harmonics
\begin{equation}
\hat{\frac{\delta T}{T}}(\vec{e}) =\sum _{\ell =2}^{+\infty }\sum
_{m=-\ell }^{m=\ell } \hat{a}_{\ell m}Y_{\ell m}(\theta ,\varphi )\, .
\end{equation}
The next step is to calculate the two-point correlation function of
temperature fluctuations. One gets
\begin{equation}
\label{corrt}
\left \langle 0 \left \vert \hat{\frac{\delta T}{T}}(\vec{e}_1)
\hat{\frac{\delta T}{T}}(\vec{e}_2)\right \vert 0\right \rangle =\sum
_{\ell =2}^{+\infty }\frac{(2\ell +1)}{4\pi }C_{\ell }P_{\ell }
\left(\cos \gamma \right)\, ,
\end{equation}
where $P_{\ell }$ is a Legendre polynomial and $\gamma $ is the angle
between the two vectors $\vec{e}_1$ and $\vec{e}_2$. The $C_{\ell }$
's are the multipole moments and have been measured with great
accuracy by the WMAP experiment~\cite{wmap}. 

\par

A remark in passing is in order at this point. As a matter of fact,
what has been measured by the WMAP satellite is the correlation
function
\begin{equation}
\left \langle \frac{\delta T}{T}(\vec{e}_1) \frac{\delta
T}{T}(\vec{e}_2)\right \rangle \, ,
\end{equation}
where the bracket denotes spatial average over the celestial sphere
and not ensemble average as in Eq.~(\ref{corrt}). Going from one to
another is not trivial and, in fact, involves profound questions which
can even go as further as problems linked to the interpretation of
Quantum Mechanics! (another related question is the problem of the
``classicalization'' of the quantum perturbations, see
Refs.~\cite{class}). In order to check that the predictions of
Eq.~(\ref{corrt}) are verified or not, one should repeat the
measurement of the CMBR map many times and see whether the result
converges toward the theoretical prediction. However, one cannot do
that because we only have at our disposal one realization, i.e. one
Universe or one CMBR map. Facing this situation, the usual strategy is
to construct an unbiased estimator of the quantity that we want to
measure (the correlation function or the multipole moments) with the
minimum possible variance so that it is very probable that the outcome
of one realization is closed to the mean
value~\cite{GM}. Unfortunately, the variance cannot be zero (in this
case only one realization would be enough to estimate the result) and
one can show that this is linked to the fact that a stochastic process
on a sphere cannot be ergodic~\cite{GM}. This variance is called the
``cosmic variance'' and is generally large on large scales. More
details on this question can be found for instance in Ref.~\cite{GM}.

\par

On large scales, i.e. for small $\ell$, one can use
Eq.~(\ref{swlarge}) to find an explicit expression of the multipole
moments. One gets
\begin{equation}
C_{\ell }=\frac{4\pi }{25}\int _0^{+\infty }\frac{{\rm d}k}{k}
j_{\ell }^2(k)k^3P_{\zeta }\, ,\quad \ell \ll 20 \, ,
\end{equation}
where $j_{\ell }$ is a spherical Bessel function of order $\ell
$. Using Eq.~(\ref{pssr}) for density perturbations (since they are
dominant) and neglecting the logarithmic corrections (which amounts to
consider that the spectrum is scale-invariant), we obtain
\begin{equation}
\label{largecl}
C_{\ell }=\frac{2H^2}{25\epsilon m_{_{\rm Pl}}^2}\frac{1}{\ell (\ell
+1)}\, ,\quad \ell \ll 20 \, .
\end{equation}
Therefore, a scale invariant spectrum implies that, on large scales,
the quantity $\ell (\ell +1)C_{\ell }$ is a constant. In order to
calculate the inflationary multipole moments $C_{\ell }$ for any $\ell
$ one must use a numerical code, for instance the CAMB
code~\cite{camb}.  Typically, one gets a plateau and then acoustic
oscillations. Here we do not treat this question but the details can
be found in Ref.~\cite{pdb}.

\par

The satellites COBE and WMAP have measured the quantity $Q/T\equiv
\sqrt{5C_2/(4\pi )}$ where $T\simeq 2.7\mbox{K}$ and have found
$Q\simeq 18\times 10^{-6}\mbox{K}$. Moreover, recent
analysis~\cite{saminf} of the WMAP data have been able to put a
constraint on the value of the slow-roll parameter $\epsilon $. It was
found that $\epsilon <0.032$. This allows us to put a constraint on
the Hubble parameter at horizon crossing. One finds
\begin{equation}
\frac{H_{_{\rm inf}}^2}{m_{_{\rm Pl}}^2}=60\pi \epsilon
\frac{Q^2}{T^2} \Rightarrow \frac{H_{_{\rm inf}}}{m_{_{\rm
Pl}}}<1.6\times 10^{-5}\, .
\end{equation}
This also puts a constraint on the amount of gravitational waves. In 
Ref.~\cite{saminf}, the following result has been obtained
\begin{equation}
\frac{C_{10}^{\rm T}}{C_{10}^{\rm S}}<0.3 \, ,
\end{equation}
that is to say the contribution of gravitational waves is already
constrained to be less than $30\%$ of the total contribution.

\par

We conclude this part by a summary of the main observational
predictions of single field inflation\index{inflation!predictions}:
(i) The universe is spatially flat: $\Omega _0=1\pm 10^{-5}$; (ii) The
spectrum of density perturbations is scale invariant
(Harrison-Zeldovich spectrum) plus logarithmic corrections which are
model dependent, i.e. $n_{_{\rm S}}=1+{\cal O}(\epsilon ,\delta )$;
(iii) There is a nearly scale invariant background of gravitational
waves, i.e. $n_{_{\rm T}}={\cal O}(\epsilon)$; (iv) The statistical
properties of the CMB anisotropies are Gaussian, i.e.  everything is
characterized by the power spectrum and we have the following
properties
\begin{equation}
\left \langle 0\left \vert \left (\hat{\frac{\delta T}{T}} \right )^3
\right \vert 0\right \rangle =0, \quad \left \langle 0\left \vert
\left (\hat{\frac{\delta T}{T}} \right )^4 \right\vert 0 \right
\rangle -3\, \left \langle 0\left \vert \left (\hat{\frac{\delta
T}{T}} \right )^2 \right\vert 0\right \rangle^2=0, \quad \mbox{etc
...}\, .
\end{equation}
This conclusion comes from the fact that the quantum state of the
perturbations is the vacuum, the ``wave function'' of which is a
Gaussian; (v) Gravitational waves are sub-dominant and there exists a
consistency check relating the importance of gravitational waves with
respect to scalar density on one hand to the tensor spectral index on
the other hand. This relation reads
\begin{equation}
\frac{C_2^{\rm T}}{C_2^{\rm S}}\simeq -f_2(h,\Omega _{\rm cdm},\Omega
_{\Lambda}, \cdots )n_{_{\rm T}} \, ,
\end{equation}
where the function $f_2$ is $f_2\simeq 5$ for the concordance model
(i.e.  the cold dark matter model plus dark energy which seems to fit
best the data at the time of writing); (vi) There are oscillations in
the power spectrum. Although this conclusion is also based on the
physics of the transfer function, the fact that the perturbations are
generated in a coherent manner plays a crucial role for the survival
of the acoustic peaks, see Ref.~\cite{dodel}.

\section{The Trans-Planckian Problem of Inflation}
\index{inflation!trans-Planckian problem}

We have seen that the CMBR anisotropies are, if the inflation theory
turns out to be correct, an observable signature of quantum
gravity. However, as it is clear from the previous considerations, the
CMBR anisotropies originate from a regime where the quantization of
the gravitational field is carried out in the standard manner. In
fact, the situation is similar to the Hawking radiation. In this last
case, we have a quantum field living in a classical background. In the
present context, we also have a field $\hat{h}_{\mu \nu}(\eta
,\vec{x})$ living in the classical FLRW Universe. Of course, the main
difference is that, in the case of inflation, the quantized test field
is the perturbed metric, i.e. is the gravitational field itself (at
least the small excitations of the gravitational field around a
classical background) contrary to the Hawking effect where the field
is just a scalar field: this is why, conceptually, the Hawking effect
does not involve quantum gravity while the theory of cosmological
perturbations does. Nevertheless, from the pure technical point of
view, we have just used the techniques of ordinary quantum field
theory in curved space-time. In this section, we suggest that the CMBR
anisotropies could also carry some signatures of quantum gravity but,
this time, originating from the non perturbative
regime~\cite{MB1}. Obviously, the price to pay is that the following
considerations are much more speculative than the rest of this review
article but the hope is to learn about quantum gravity, maybe in the
non-linear regime. Therefore, it seems that the potential reward is
worth the speculation.

\par

The inflationary trans-Planckian issue is based on a very simple
remark~\cite{MB1}. If we assume a model, for instance a potential of
the type given by Eq.~(\ref{pot}) (here, we choose $n=4$ to be
concrete), then one can calculate the coupling constant $\lambda
_n$. For this purpose, it is convenient to express everything in terms
of $N_*$, the number of e-folds before the end of inflation at which
the modes crossed out the Hubble radius, see
Eq.~(\ref{efoldexit}). The corresponding value of the inflaton field is
given by $\varphi _*^2=m_{_{\rm Pl}}^2(N_*+1)/\pi $. Therefore, the
Hubble parameter can be expressed as $H_*^2=\lambda _4m_{_{\rm
Pl}}^2(\varphi _*/m_{_{\rm Pl}})^4=\lambda _4 m_{_{\rm
Pl}}^2(N_*+1)^2/\pi ^2$. Finally, since the slow-roll parameter
$\epsilon $ is given by $\epsilon =(N_*+1)^{-1}$, one arrives at
\begin{equation}
\frac{H_*^2}{\epsilon m_{_{\rm Pl}}^2}=\frac{1}{\pi ^2}\lambda
_4(N_*+1)^3\, .
\end{equation}
The scale of inflation only enters the above equation through $N_*$
and the corresponding dependence is logarithmic, see
Eq.~(\ref{efoldexit}) hence very mild. One can thus use this formula
to determine the coupling constant almost independently of $H_{_{\rm
inf}}$. Using Eq.~(\ref{largecl}) for $\ell =2$ and the link between
$Q$ and $C_2$, one finds that $\lambda _4\simeq 10^{-13}$, where we
have used $N_*\simeq 60$. As already mentioned, this means that the
total number of e-folds is huge, $N_{_{\rm T}}\simeq 4.9\times
10^8$. As a result, the Hubble radius today, $\ell _{_{\rm
H}}=10^{61}\ell _{_{\rm Pl}}$ ($h=0.5$), where $\ell _{_{\rm Pl}}$ is
the Planck length, was equal to $\simeq {\rm e}^{-10^8}\ell _{_{\rm
Pl}}\simeq 10^{-4.7\times 10^7}\ell _{_{\rm Pl}}$ at the beginning of
inflation, i.e., very well below the Planck length!

\par

One can view the problem differently and ask how many e-folds before
the end of inflation a given scale was equal to the Planck length. The
answer can be easily calculated from Eq.~(\ref{efoldexit}) and reads
\begin{eqnarray}
N_{_{\rm Pl}}(\lambda ) &=& N_*(\lambda )- \log _{10}\left(\frac{H_{\rm
inf}}{m_{_{\rm Pl}}}\right)\times \ln 10 
\\
&\simeq & \ln \left(\frac{\lambda }{\ell _{_{\rm H}}}\right)
+\left[29- \log _{10}\left(\frac{T_{_{\rm RH}}}{m_{_{\rm
Pl}}}\right)\right]\times \ln 10\, .
\end{eqnarray} 
If one takes the fiducial values $H_{_{\rm inf}}\simeq
10^{14}\mbox{GeV}$, $T_{_{\rm RH}}=M_{_{\rm inf}}\simeq
10^{16.5}\mbox{GeV}$, one finds that the Planckian region was reached
only $11$ e-folds before the modes crossed out the horizon during
inflation, see Fig.~\ref{fig:scales}. For instance, for the mode
$\lambda =\ell _{_{\rm H}}$, this means $70 $ e-folds before the end
of inflation. Of course, if the scale of inflation is smaller, then
the number of e-folds before the exit of the Planckian region and the
exit of the horizon can be bigger.

\par

The following point should also be emphasized. At the time at which
the modes of astrophysical interest today exit the Planckian region,
the value of the Hubble parameter is generically well-below the
Planckian mass. This means that the use of a classical FLRW background
is well justified. The trans-Planckian problem concerns only the
fluctuations and has to do with the fine structure of the Universe or
with the ``Planckian foam'' but does necessitate a full quantum
gravity description of the evolution of the underlying manifold (for
instance, one does not need quantum cosmology).

\par

Having in mind the above considerations, the trans-Planckian problem
of inflation consists in the following~\cite{MB1}. It is likely that
the framework of standard quantum field theory described in the
previous section and used in order to establish what the predictions
of inflation are breaks down when the modes under consideration have a
wavelength smaller than the Planck length. Therefore, there is the
danger that the so far successful predictions of inflation are in fact
based on a theory used outside its domain of validity. In other words,
there is the problem that the predictions of inflation could in fact
depend on physics on scales shorter than the Planck length, a physics
which is clearly largely unknown.

\par

Is it really so? In trying to answer this question we immediately face
the problem that the trans-Planckian physics is presently unknown and
that, as a consequence, it is a priori impossible to study its
influence on the inflationary predictions. To circumvent this
difficulty, one studies the robustness of inflationary predictions to
ad-hoc (``reasonable'') changes in the standard quantum field theory
framework supposed to mimic the modifications caused by the actual
theory of quantum gravity. If the predictions are robust to some
reasonable changes, then there is the hope that they will be robust to
the modifications induced by the true theory of quantum gravity. On
the other hand, if the predictions are not robust, the knowledge of
the exact theory seems to be required in order to predict exactly what
the changes are. The next question is of course which kind of
modifications can we introduce in the theory in order to test its
robustness? Many proposals have been made and discussed recently in
the literature~\cite{MB1,trans,LLMU,BM3}. Here, we concentrate on two
possibilities: the modified dispersion relation and the so-called
``minimal'' approach.

\subsection{Modified Dispersion Relations}
\index{modified dispersion relations}

Let us start with the modified dispersion relations. The term $k^2$ in
Eq.~(\ref{parameq}) originates from the use of the standard dispersion
relation $\omega _{_{\rm phys}}=k_{_{\rm phys}}$. In condensed matter
physics, it is known that the dispersion relation starts departing
from the linear relation $\omega =k$ on scales of the order of the
atomic separation: the mode feels the granular nature of matter. In
the same way, one can expect the dispersion relation to change when
the mode starts feeling the discreteness of space-time on scales of
the order of the Planck (string) length. Therefore, our method is to
replace the linear dispersion relation $\omega _{_{\rm phys}}=k_{_{\rm
phys}}$ by a non standard dispersion relation $\omega _{_{\rm
phys}}=\omega _{_{\rm phys}}(k)$, this non linear relation having of
course the property that $\omega _{_{\rm phys}}\simeq k_{_{\rm phys}}$
for $k\ll k_{_{\rm C}}$ where $k_{_{\rm C}}$ is a new scale introduced
in the theory which could be, for instance the string scale. In the
context of cosmology, this amounts to replacing the square of the
comoving wavenumber $k^2$ with
\begin{equation}
k^2 \, \rightarrow \, k_{\rm eff}^2(k,\eta ) \equiv 
a^2(\eta )\omega _{_{\rm phys}}^2\biggl[\frac{k}{a(\eta )}\biggr]\, .
\end{equation}
Therefore, this implies that we now deal with a time-dependent
dispersion relation, a result first obtained in Ref.~\cite{MB1}. As a
consequence, the equation of motion~(\ref{parameq}) now takes the form
\begin{equation} 
\label{eom2}
\mu '' + \left[k_{\rm eff}^2(k,\eta ) - {{z''} 
\over z}\right]\mu = 0 \, .
\end{equation}
The effect of the new physics is to change the time-dependent
frequency $\omega (k,\eta )$ of the parametric oscillator. Let us
remark that a more rigorous derivation of this equation, based on a
variational principle, has been provided in Ref.~\cite{LLMU}.

\par

Then, the only question is whether the fact that we now have a new
time-dependent frequency can modify the spectrum $k^3\vert \mu \vert
^2$ or not? As we now demonstrate, this depends on whether the
evolution of the modes is adiabatic or not in the trans-Planckian
region. Indeed, if the dynamics is adiabatic throughout (in particular
if the $z''/z$ term is negligible), the WKB approximation holds and
the solution is always given by
\begin{equation} 
\label{WKBsol}
\mu (\eta )\, \simeq \, \frac{1}{\sqrt{2k_{\rm eff}(k,\eta )}}
\exp\left[-i\int _{\eta _{\rm ini}}^{\eta }k_{\rm eff}(k,\tau ){\rm
d}\tau \right] \, ,
\end{equation} 
where $\eta_{\rm ini}$ is some initial time. Therefore, if we start
with a positive frequency solution only and uses this solution, one
finds that no negative frequency solution appears. Deep in the region
where $k_{\rm eff} \simeq k$, i.e. for $k\ll k_{_{\rm C}}$, the
solution becomes
\begin{equation}
\mu (\eta ) \simeq {1 \over {\sqrt{2k}}} \exp\left[-ik\eta -i \int
_{\eta _{\rm ini}}^{\eta _1}k_{\rm eff}(k,\tau ) {\rm d}\tau \right]\,
,
\end{equation}
where $\eta _1$ is the time at which $k_{\rm eff}\simeq k$. Up to an
``accumulated'' phase which will disappear when we calculate the
modulus $\vert \mu \vert ^2$, we recover the standard vacuum solution
${\rm e}^{-ik\eta }/\sqrt{2k}$ and hence the standard spectrum. We
have thus identified the criterion which controls whether the spectrum
will be changed or not: in order to get a modification, the dispersion
relation in the trans-Planckian region must be such that the WKB
approximation is violated. This constrains the shape of the modified
dispersion relation. It is possible to give the conditions for
violation of the WKB approximation. Given an equation of the form $\mu
''+\omega ^2\mu =0$ (in the present context, one has $\omega
^2=k^2_{\rm eff}-z''/z$), the WKB approximation is valid if the
following quantity is small in the trans-Planckian region~\cite{wkb}
\begin{equation}
\left \vert \frac{Q}{\omega ^2}\right \vert \ll 1\, ,
\end{equation}
where $Q$ is defined by the following expression $Q=3(\omega
')^2/(4\omega ^2)-\omega ''/(2\omega )$. Then, one can insert in the
previous expression one's favorite dispersion relation ans see whether
this leads to a new spectrum. This has been done recently in the
literature, see Refs.~\cite{trans}. For instance, one can show that
the dispersion relations introduced in Refs.~\cite{Unruh,Jacobson} do
not lead to any modification. An example where modifications are
present has been studied in Ref.~\cite{LLMU}. However, it remains to
be studied whether this can be made compatible with other studies on
the subject, in particular those using astrophysical observations to
constraint the deviations from the law $\omega =k$~\cite{ted}. Rather
than studying these examples in great details, we now turn to a new
way of modeling the trans-Planckian regime.

\subsection{The Minimal Approach}

Modifying the dispersion relation is equivalent to changing the form
of the equation of motion for the perturbations. The minimal approach
consists in working with the same equation of motion (with a standard
dispersion relation hence the name ``minimal approach'') but with
modified initial conditions. For a given Fourier mode, the initial
conditions are fixed when the mode emerges from the trans-Planckian
region, i.e. when its wavelength becomes equal to a new fundamental
characteristic scale $\ell_{_\mathrm{C}}=1/k_{_{\rm C}}$. The time
$\eta _k$ of mode ``appearance'' with comoving wavenumber $k$, can be
computed from the condition
\begin{equation}
\lambda (\eta _k)=\frac{2\pi }{k}a(\eta _k)=\ell _{_{\rm C}}
\equiv \frac{2\pi }{M_{_{\rm C}}} ,
\end{equation}
which implies that $\eta _k$ is a function of $k$. This has to be
compared with the standard inflationary calculations where the initial
time is taken to be $\eta _k=-\infty $ for any Fourier mode $k$ and
where, in a certain sense, the initial time does not depend on
$k$. Then, a crucial question is in which state the Fourier mode is
created at the time $\eta _k$ (here, we cannot take the limit $k\eta
\to -\infty$ anymore). The only requirement is that, if we send the
new scale $M_{_{\rm C}}$ to infinity (i.e. if there is no
trans-Planckian region), then one must recover the standard WKB
vacuum. Therefore, the most general parametrization of these initial
conditions read
\begin{eqnarray}
\label{ci1}
\mu(\eta _k) =\mp
\frac{c_k+d_k}{\sqrt{2\omega _{_{\rm S,T}}(\eta _k)}}
\frac{4\sqrt{\pi }}{m_{_{\rm Pl}}} \, ,
\quad 
\label{ci2}
\mu'(\eta _k) =
\pm i\sqrt{\frac{\omega _{_{\rm S,T}}(\eta _k)}{2}}
\frac{4\sqrt{\pi }(c_k-d_k)}{m_{_{\rm Pl}}} .
\end{eqnarray}
where the coefficients $c_k$ and $d_k$ are {\it a priori} two
arbitrary complex numbers satisfying the condition $\vert c_k\vert
^2-\vert d_k\vert ^2=1$ and which can be expanded as
\begin{equation}
c_k=1+y\sigma_0+\cdots \, \quad d_k=x\sigma_0+\cdots \, ,
\end{equation}
where $\sigma_0\equiv H/M_{_{\rm C}}$. When $M_{_{\rm C}}$ is sent to
infinity then $\sigma _0\to 0$, $c_k=1$, $d_k=0$ and, indeed, we
recover the standard vacuum.  Since there are two energy scales in the
problem, namely the Hubble parameter $H$ during inflation and the new
scale $M_{_{\rm C}}$, it is natural that the final result is expressed
in terms of their ratio $H/M_{_{\rm C}}$, which is typically a small
parameter. The parameters $x$ and $y$ are considered as free
parameters that are not fixed by any existing well-established
theories except, as already mentioned above, that they should be such
that the relation $\vert c_k\vert ^2-\vert d_k\vert ^2=1$ is
satisfied. One easily shows that this implies $y+y^*=0$ at leading
order in $\sigma _0$. Expanding everything in terms of $\sigma_0$, one
arrives at~\cite{BM3}
\begin{eqnarray}
\label{pssrs2}
& & k^3P_{\zeta }=\frac{H^2}{\pi \epsilon m_{_{\rm Pl}}^2}
\biggl\{1-2(C+1)\epsilon -2C(\epsilon -\delta )-2(2\epsilon -\delta )
\ln \frac{k}{k_*}-2\vert x\vert \sigma _0
\nonumber \\
& & \times 
\biggl[1 -2(C+1)\epsilon -2C(\epsilon -\delta )
-  2(2\epsilon -\delta )\ln \frac{k}{k_*}\biggr]
\times \cos \biggl[\frac{2}{\sigma _0}
\biggl(1+\epsilon 
\nonumber \\
& &  +\epsilon \ln \frac{k}{a_0 M_{_{\rm C}}}\biggr)
+\varphi \biggr]
-2\vert x\vert \sigma _0\pi (2\epsilon -\delta )
\sin \biggl[\frac{2}{\sigma _0}
\biggl(1+\epsilon +\epsilon \ln \frac{k}{a_0 M_{_{\rm C}}}\biggr)
+\varphi \biggr]
\biggr\},
\nonumber \\
\label{pst}
& & k^3P_h = \frac{16 H^2}{\pi m_{_{\rm Pl}}^2}
\biggl\{1-2(C+1)\epsilon -2\epsilon \ln \frac{k}{k_*}
-2\vert x\vert \sigma _0\biggl[1-2(C+1)\epsilon -2\epsilon 
\ln \frac{k}{k_*}\biggr]
\nonumber \\
& & \times
\cos \biggl[\frac{2}{\sigma _0}
\biggl(1+\epsilon +\epsilon \ln 
\frac{k}{a_0 M_{_{\rm C}}}\biggr)+\varphi \biggr]
-2\vert x\vert \sigma _0\pi\epsilon
\sin \biggl[\frac{2}{\sigma _0}
\biggl(1+\epsilon 
\nonumber \\
& & +\epsilon \ln \frac{k}{a_0M_{_{\rm C}}}\biggr)
+\varphi \biggr]
\biggr\}\, ,
\end{eqnarray}
where $\varphi $ is the argument of the complex number $x$, i.e
$x\equiv \vert x\vert {\rm e}^{i\varphi }$. These expressions should
be compared with Eqs.~(\ref{pssr}) and~(\ref{pssrgw}). The effect of
the trans-Planckian corrections is clear: superimposed oscillations in
the power spectra have appeared. The magnitude of the trans-Planckian
corrections are linear in the parameter $\sigma _0$ and their
amplitude is given by $\vert x\vert \sigma _0$. The wavelength of the
oscillations can be expressed as $\Delta k/k=\sigma_0\pi/\epsilon$.

\par

\begin{figure}
\includegraphics[angle=0,width=10.8cm]{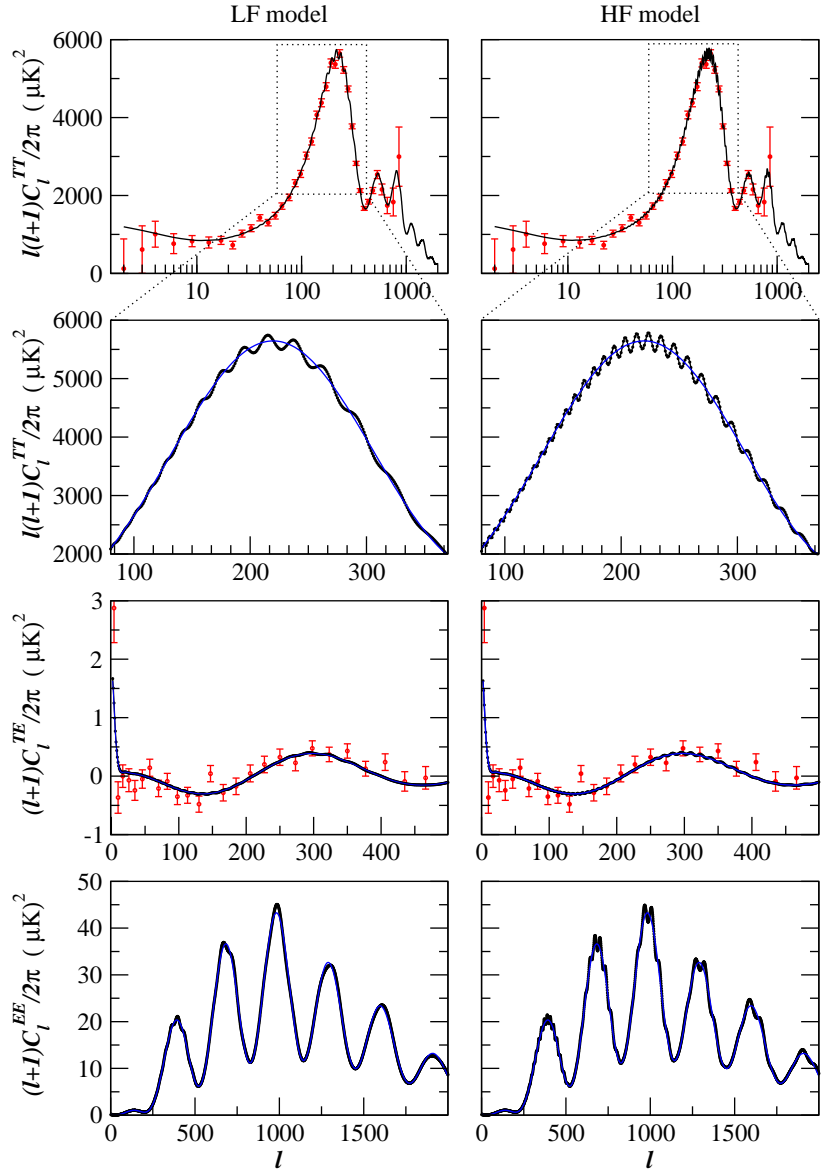}
\caption{Angular TT, TE and EE power spectra for two different
trans-Planckian models, one with low frequency (LF) superimposed
oscillations, the other with high frequency (HF) oscillations, for
details see Refs.~\cite{MR1,MR2}. A zoom of the temperature multipole
moments in the first Doppler peak region is also shown (black curve)
and compared with the standard slow-roll prediction (blue curve)
calculated with the same cosmological parameters.}
\label{fig:cls}
\end{figure}

The above calculation provides us with an explicit example where the
observational predictions of inflation are modified by the
trans-Planckian physics. Let us now study this question in more
details. Using Eq.~(\ref{swlarge}), one can evaluate the modifications
of the multipoles moments caused by the trans-Planckian
corrections. In the limit $\epsilon/\sigma_0\gg \ell$, one
gets~\cite{MR1,MR2}
\begin{eqnarray}
\label{eq:largescalestpl}
\ell(\ell +1) C_\ell & \simeq & \frac{2 H^2}{25 \epsilon m_{_{\rm
Pl}}^2} (1-2\epsilon) \biggl\{ 1 + \sqrt{\pi} \frac{|x|\sigma_0
\ell(\ell+1)}{\left(\epsilon/\sigma_0\right)^{5/2}} \nonumber \\
& & \times \cos\left[\pi \ell + \frac{2}{\sigma_0}\left(1 +
\epsilon\ln \frac{\epsilon/\sigma_0}{a_0 M_{_{\rm C}} r_{\rm lss}}
\right) + \varphi - \frac{\pi}{4}\right] \biggr\}.
\end{eqnarray}
This expression should be compared with Eq.~(\ref{largecl}). The
oscillations in the power spectra are transfered to the multipole
moments, at least at relatively small scales. At large $\ell $, or for
not too small values of $\sigma_0$, the above equation quickly becomes
invalid and an accurate estimation can be made only with the help of
numerical calculations. The result in plotted in Fig.~\ref{fig:cls}
for the temperature fluctuations but also for the polarization, for
details see Ref.~\cite{MR1,MR2}.
In those references, a detailed comparison of the trans-Planckian
signal with the recently released high accuracy WMAP data has been
performed. The main result is that, with the oscillations taken into
account, it is possible to decrease the $\chi ^2$ significantly.
Instead of $\chi ^2\simeq 1431$ for $1342$ degrees of freedom for the
standard slow-roll power spectra, one now obtains $\chi^2\simeq 1420$
for $1340$ degrees of freedom, i.e. $\Delta \chi ^2\simeq 10$ compared
to WMAP one. The reason for such an important improvement of the $\chi
^2$ is due to the presence of the oscillations which permit a better
fit of the cosmic variance outliers at small scales. The main question
is of course the statistical significance of this result. In
Ref.~\cite{MR1,MR2}, the so-called F-test has been used and indicates
that the result is significant. However, it is clear that other
statistical tests, a complete exploration of the parameter space and,
of course, new data, should be used before one can really conclude
that superimposed oscillations are really present in the CMBR
multipole moments. A fair description of the present situation is that
there seems to be a hint for an interesting feature in the CMBR data
and that, maybe, this feature is a signature of very high energy
physics (it is clear that the oscillations, if their presence is
confirmed, could have another physical origin).

\par

Finally, we would like to conclude by a comment on the back-reaction
problem. This question is crucial for the consistency of the approach
used before. It is clear that the energy density of the perturbations
must be smaller or equal than that of the inflationary
background. This leads to the condition $\vert x\vert \le \sqrt{3\pi}
m_{_{\rm Pl}}/M_{_{\rm C}}$ which amounts to
\begin{equation}
\label{eq:brvalid}
\vert x\vert \sigma _0\le 10^4 \times 
\frac{\sigma _0^2}{\sqrt{\epsilon }} \, .
\end{equation}
It is important to emphasize that the above constraint is only a
sufficient condition, but by no means, unless proved otherwise, a
necessary condition. In general, this constraint is difficult to
satisfy. Some of the best fits described above suffer from this
back-reaction problem, see Ref.~\cite{LLMU,MR1,MR2}. In fact, the
above formula expresses a generic difficulty of the trans-Planckian
question, this difficulty being present regardless of the approach
used in order to model the new physics. The presence of
trans-Planckian corrections means the presence of particles (with
respect to the standard vacuum) the energy density of which is very
easily of the order of the background energy density. On the other
hand, if we try to satisfy the back-reaction constraint then the
signal very easily becomes tiny and, hence, non observable. A major
advance, which would allow us to escape the previous vicious circle,
would be to calculate explicitly the effect of the
back-reaction. Unfortunately, for the moment, this is still an open
question and more work is required to tackle this very important task.

\section*{Acknowledgments}

It is a pleasure to thank G.~Amelino-Camelia and J.~Kowalski-Glikman
for inviting me to lecture at this school and for their hospitality. I
am very grateful to all the participants for very interesting
discussions. I also thank P.~Brax and M.~Lemoine for careful reading
of the manuscript.

%
%
%


%
%

%

\begin{thebibliography}{99.}
%
%
%

\bibitem{guth} A.~Guth, Phys. Rev. D {\bf 23}, 347 (1981).

\bibitem{inflation}
A.~Linde, Phys. Lett. B {\bf 108}, 389 (1982); A.~Albrecht and
P.~J.~Steinhardt, Phys. Rev. Lett. {\bf 48}, 1220 (1982); A.~Linde,
Phys. Lett. B {\bf 129}, 177 (1983).

\bibitem{procbrazil} J.~Martin, Proceedings of the XXIV Brazilian
National Meeting on Particles and Fields, Caxambu, Brazil, (2004),
{\tt astro-ph/0312492}.

\bibitem{LR} D.~Lyth and A.~Riotto, Phys. Rept. {\bf 314}, 
1 (1999), {\tt hep-ph/9807278}.

\bibitem{Linde} A.~Linde, Talk at the Nobel Symposium ``Cosmology 
and String Theory'', (2003), {\tt hep-th/0402051}.

\bibitem{turner} M.~Turner, Phys. Rev. D {\bf 28}, 1243 (1983).

\bibitem{preheating} L.~Kofman, A.~Linde and A.~Starobinsky,
Phys. Rev. D {\bf 56}, 3258 (1997), {\tt hep-ph/9704452}.

\bibitem{MFB} V.~F.~Mukhanov, H.~A.~Feldman, and R.~H.~Brandenberger,
Phys. Rep. {\bf 215}, 203 (1992).

\bibitem{stewart} J.~Stewart, Class. Quantum Grav. {\bf 7}, 1169
(1990).

\bibitem{Bardeen} J.~A.~Bardeen, Phys. Rev. D {\bf 22}, 1882 (1980).

\bibitem{MS1} J.~Martin and D.~J.~Schwarz, Phys. Rev. D {\bf 57}, 3302
(1998), {\tt gr-qc/970449}.

\bibitem{RK} R.~Kerner, Gen. Rel. Grav. {\bf 9}, 257 (1978).

\bibitem{L} D.~H.~Lyth, Phys. Rev. D {\bf 31}, 1792 (1985).

\bibitem{Gri} L.~P.~Grishchuk, Zh. Eksp. Teor. Fiz {\bf 67}, 825 (1974).

\bibitem{cobe} G.~F.~Smooth, {\it et al.}, Astrophys. J. {\bf 396}, L1
(1992).

\bibitem{pdb} P.~de~Bernardis, this volume.

\bibitem{sw} R.~K.~Sachs and A.~M.~Wolfe, Astrophys. J. {\bf 147}, 73
(1967).

\bibitem{panek} M.~Panek, Phys. Rev. D {\bf 49}, 648 (1986).

\bibitem{cohen} C.~Cohen-Tannoudji, J.~Dupont-Roc and G.~Grynberg,
``Photons et atomes, Introduction a l'\'electrodynamique quantique'',
Editions du CNRS, (1987).

\bibitem{KT} E.~Kolb and M.~Turner, ``The Early Universe'', Frontier
in Physics, Perseus Publishing, (1990).

\bibitem{MS2} J.~Martin and D.~J.~Schwarz, Phys.~Rev.~D {\bf 62},
103520 (2000), {\tt astro-ph/9911225}; J.~Martin, A.~Riazuelo and
D.~J.~Schwarz, Astrophys.~J.~{\bf 543}, L99 (2000), {\tt
astro-ph/0006392}.

\bibitem{wmap} C.~L.~Bennet {\it et al.},
Astrophys. J. Suppl. \textbf{148}, 1 (2003), {\tt astro-ph/0302207};
G.~Hinshaw {\it et al.}, Astrophys. J. Suppl. \textbf{148}, 135
(2003), {\tt astro-ph/0302217}; L.~Verde {\it et al.},
Astrophys. J. Suppl. \textbf{148}, 195 (2003), {\tt astro-ph/0302218};
H.~V.~ Peiris {\it et al.}, Astrophys. J. Suppl. \textbf{148}, 213
(2003), {\tt astro-ph/0302225}.

\bibitem{class} A.~Albrecht, P.~Ferreira, M.~Joyce and T.~Prokopec,
Phys. Rev. D {\bf 50}, 4807 (1994); D.~Polarski and A.~Starobinsky, 
Class. Quantum Grav. {\bf 13}, 377 (1996).

\bibitem{GM} L.~P.~Grishchuk and J.~Martin, Phys. Rev. D {\bf 56},
1924 (1997), {\tt astro-ph/9702018}.

\bibitem{camb} A.~Lewis, A.~Challinor and A.~Lasenby,
Astrophys. J. {\bf 538}, 473 (2000), {\tt astro-ph/9911177},
{\tt http://camb.info}; S.~Leach,
{\tt http://astronomy.sussex.ac.uk/~sleach/inflation/camb-inflation.html}.

\bibitem{saminf} S.~Leach and A.~Liddle, Phys. Rev. {\bf D 68},
123508 (2003), {\tt astro-ph/0306305}.

\bibitem{dodel} S.~Dodelson, {\tt hep-ph/0309057}.

\bibitem{MB1} J.~Martin and R.~H.~Brandenberger, Phys. Rev. {\bf D
63}, 123501 (2001), {\tt hep-th/0005209}; R.~H.~Brandenberger and
J.~Martin, Mod. Phys. Lett. {\bf A 16}, 999 (2001), {\tt
astro-ph/0005432}.

\bibitem{trans} J.~C.~Niemeyer, Phys. Rev. {\bf D 63}, 123502 (2001),
{\tt astro-ph/0005533}; A.~Kempf, Phys. Rev. {\bf D 63}, 083514
(2001), {\tt astro-ph/0009209}; R.~Easther, B.~R.~Greene, W.~H.~Kinney
and G.~Shiu, Phys. Rev. {\bf D 64}, 103502 (2001), {\tt
hep-th/0104102}; R.~Easther, B.~R.~Greene, W.~H.~Kinney and G.~Shiu,
Phys. Rev.  {\bf D 67}, 063508 (2003), {\tt hep-th/0110226}; A.~Kempf
and J.~C.~Niemeyer, Phys. Rev. {\bf D 64}, 103501 (2001), {\tt
astro-ph/0103225}; R.~H.~Brandenberger and P.~M.~Ho, Phys. Rev. {\bf D
66}, 023517 (2002), {\tt hep-th/0203119}; S.~F.~Hassan and
M.~S.~Sloth, {\tt hep-th/0204110}; F.~Lizzi, G.~Mangano, G.~Miele and
M.~Peloso, JHEP {\bf 0206}, 049 (2002), {\tt hep-th/0203099};
U.~H.~Danielsson, Phys. Rev. {\bf D 66}, 023511 (2002), {\tt
hep-th/0203198}; R.~Easther, B.~R.~Greene, W.~H.~Kinney and G.~Shiu,
Phys. Rev.  {\bf D 66}, 023518 (2002), {\tt hep-th/0204129};
G.~L.~Alberghi, R.~Casadio and A. Tronconi, {\tt gr-qc/0303035};
C.~Armend\'ariz-Pic\'on and E.~A.~Lim, {\tt hep-th/0303103}.

\bibitem{LLMU} M.~Lemoine, M.~Lubo, J.~Martin and J.~P.~Uzan,
Phys.~Rev.~{\bf D 65}, 023510 (2002), {\tt hep-th/0109128}.

\bibitem{BM3}
J.~Martin and R.~H.~Brandenberger, Phys. Rev. \textbf{D 68}, 063513
(2003), {\tt hep-th/0305161}.

\bibitem{wkb} J.~Martin and D.~J.~Schwarz, Phys.~Rev.~D {\bf 67},
083512 (2003), {\tt astro-ph/0210090}.

\bibitem{Unruh} W.~Unruh, Phys. Rev. {\bf D 51}, 2827 (1995).

\bibitem{Jacobson} S.~Corley and T.~Jacobson, Phys. Rev. {\bf D 54},
1568 (1996), {\tt hep-th/9601073}; S.~Corley, Phys. Rev. {\bf D 57},
6280 (1998), {\tt hep-th/9710075}.

\bibitem{ted} T.~Jacobson, this volume.

\bibitem{MR1} J.~Martin and C.~Ringeval, Phys. Rev. {\bf D 69}, 064406
(2004), {\tt astro-ph/0310382}.

\bibitem{MR2} J.~Martin and C.~Ringeval, {\tt astro-ph/0402609}.

\end{thebibliography}
%



\printindex
\end{document}